\documentclass[11pt]{article}

\usepackage{graphicx,ulem,cite} 
\usepackage{color}

\input babsym

\newcommand{\gfrac}[2]{\displaystyle\frac{#1}{#2}}
\newcommand{\dd}{\mbox{d}}

\newcommand{\Black}[1]{\textcolor{black}{#1}}
\newcommand{\Blue}[1]{\textcolor{blue}{#1}}
\newcommand{\Red}[1]{\textcolor{red}{#1}}

\newcommand{\Green}[1]{\textcolor{green}{#1}}

\newcommand{\Magenta}[1]{\textcolor{magenta}{#1}}

\title{Low-energy hadronic cross sections measurements at \babar, and implication for the $g-2$ of the muon} 

\author{Denis Bernard, 
\\
~
\\
LLR, Ecole Polytechnique, CNRS/IN2P3, 91128 Palaiseau, France
\\
~
\\
On behalf of the \babar\ Collaboration
}

\begin{document} 

\maketitle 

\begin{abstract}
The \babar\ Collaboration has an intensive program studying the cross
sections of hadron production in low-energy \epem\ annihilation,
accessible {\sl via} initial-state radiation.  Our measurements allow
a significant improvement in the precision of the predicted value of
the muon anomalous magnetic moment.  These improvements are necessary
for shedding light on the current $> 3$ sigma difference between the
predicted and the experimental values.  We have published results on a
number of processes with two to six hadrons in the final state, and
other final states are currently under investigation. We report here
on the most recent results obtained by analysing the entire \babar\
dataset.

\end{abstract}

~ 

~ 

~ 

~ 

~ 

~ 

\begin{center}
\large \textbf{DIS2016,
24th workshop on Deep-Inelastic Scattering and Related Subjects. 
\\
11-15 April 2016, DESY Hamburg
}
\end{center}

\section{The muon gyromagnetic factor and ``anomalous'' moment}

As a result of more than three decades of intense efforts to validate
every corner of the standard model (SM) of elementary particles and
their interactions, and to submit it to a redundant metrology with an
always increasing precision, the SM has only become
more and more ``standard'', with some very few exceptions that include
the ``tension'' between the theoretical prediction and the unique
precise experimental measurement of the ``anomalous'' magnetic moment
of the muon, $a_\mu$, which is the relative deviation of the
gyromagnetic factor, $g_\mu$, from the value of $g=2$ for a pointlike
Dirac particle, i.e. $a_\mu \equiv (g_\mu -2)/2$.

\section{$a_\mu$: predictions and measurement}

\begin{figure}[Htb]
\begin{center} \small 
\begin{tabular}{ccccccc}
$1^{rst}$ & $2^{nd}$ & $3^{rd}$ & $4^{th}$ & $5^{th}$ \\
\includegraphics[width=0.16\linewidth]{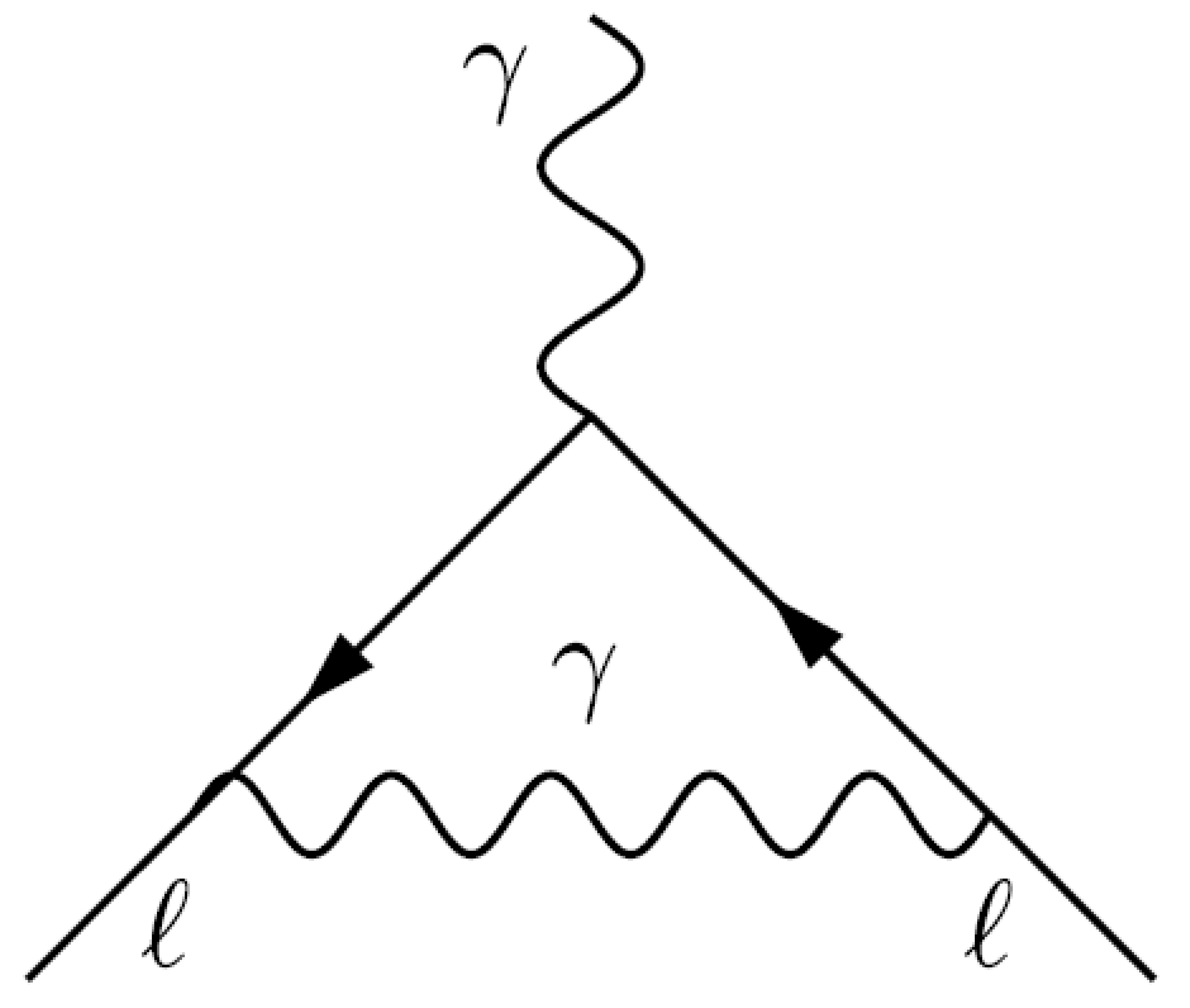}
&
\includegraphics[width=0.16\linewidth]{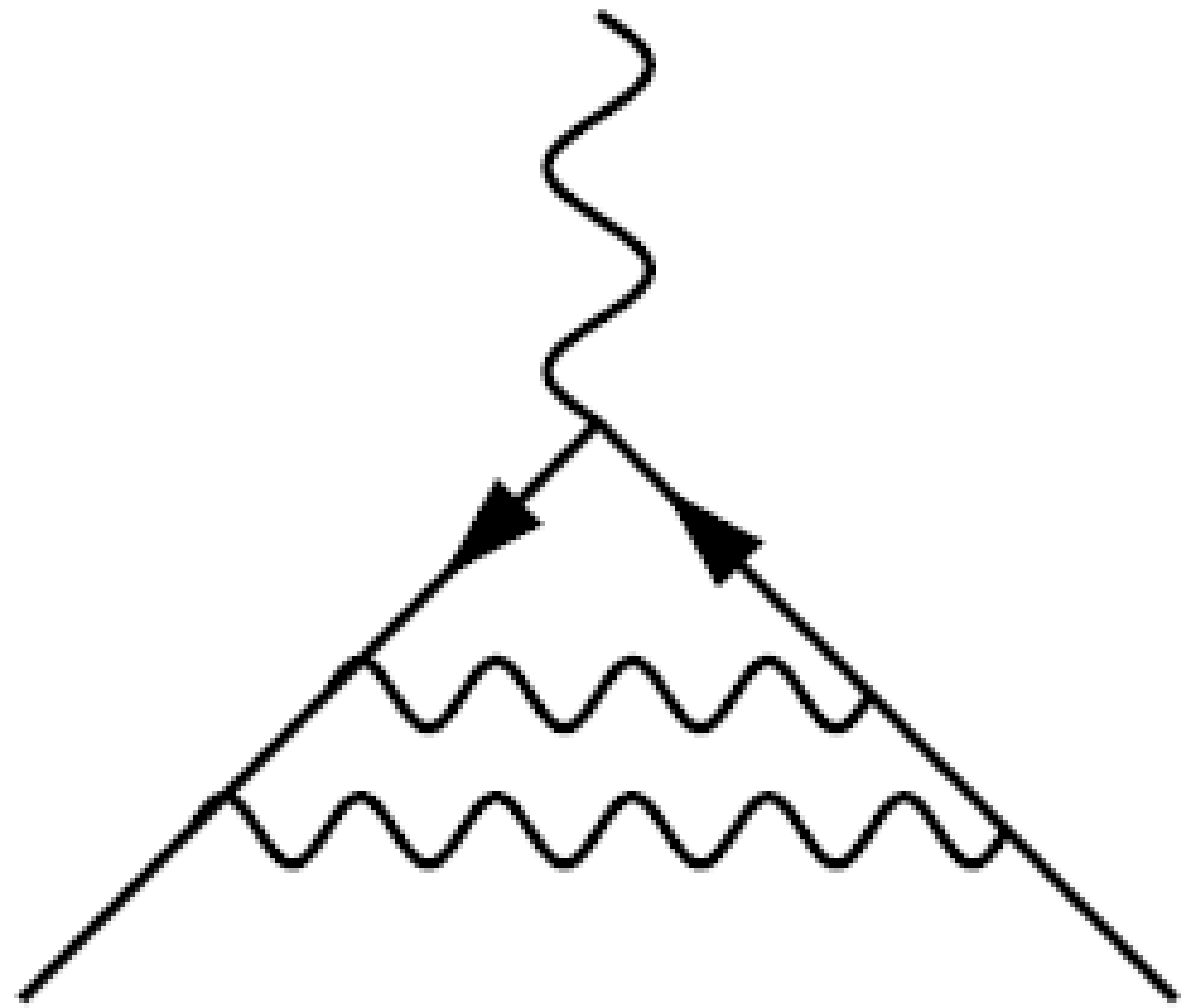}
&
\includegraphics[width=0.08\linewidth]{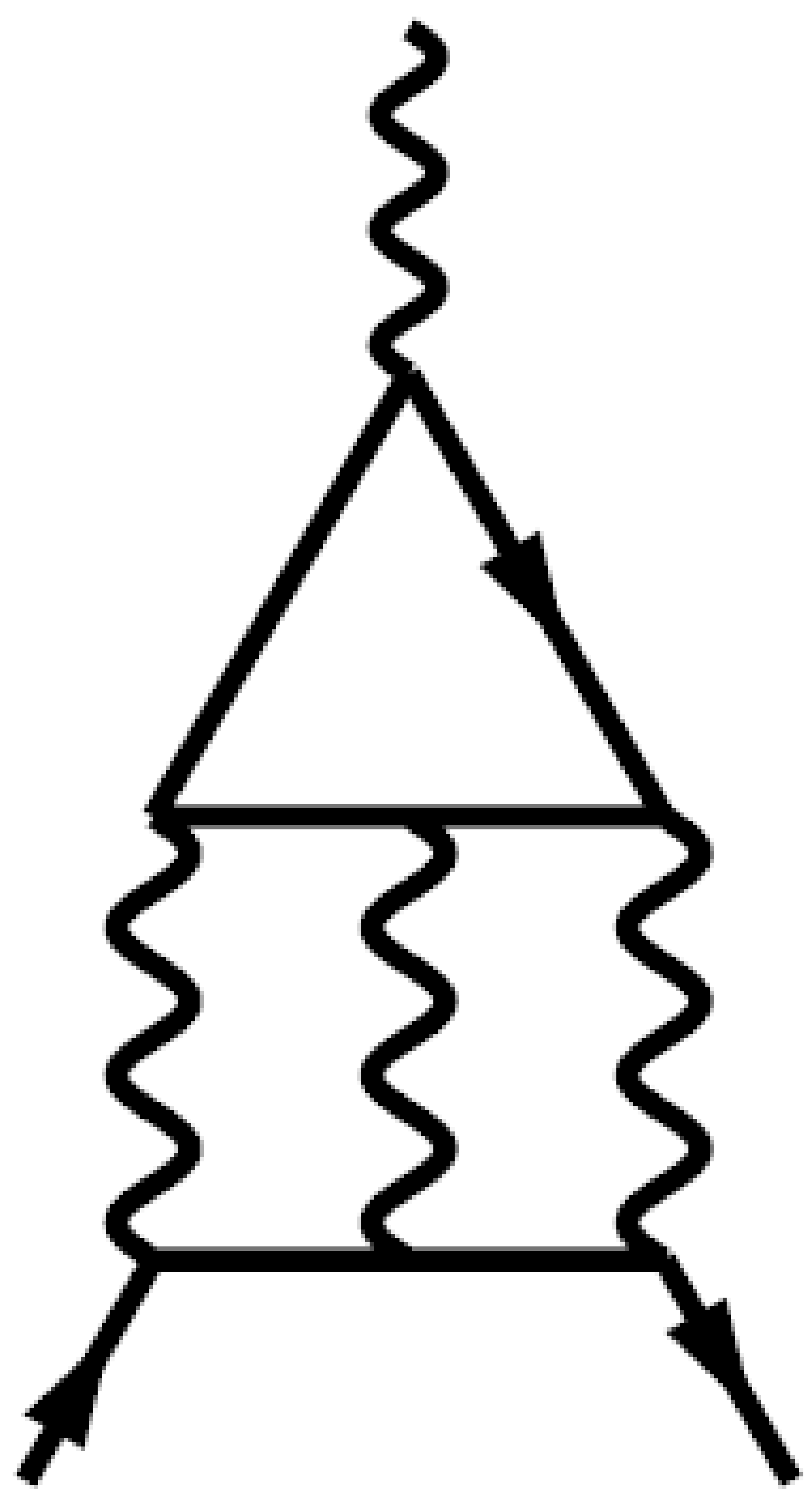}
&
\includegraphics[width=0.13\linewidth]{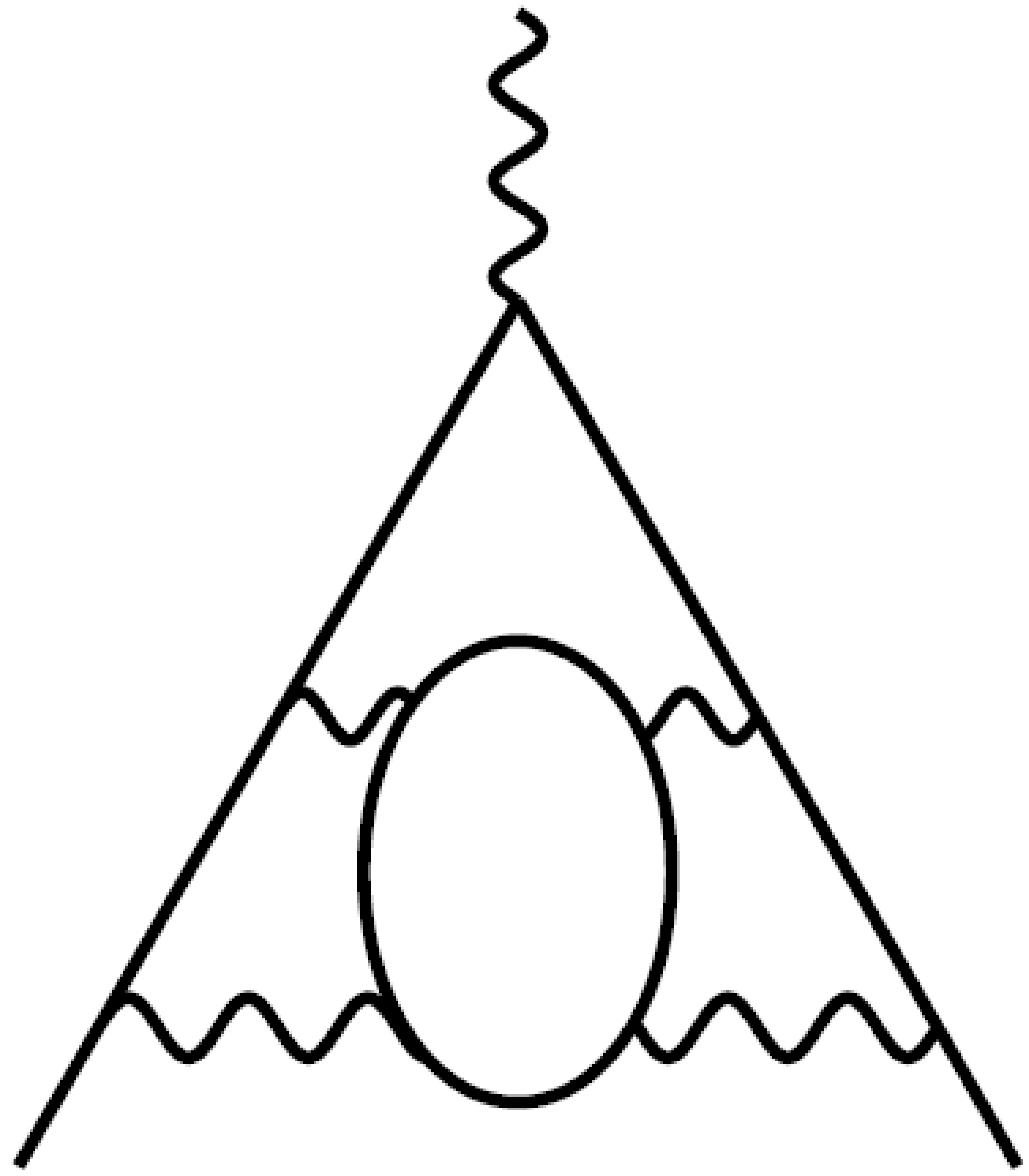}
&
\includegraphics[width=0.08\linewidth]{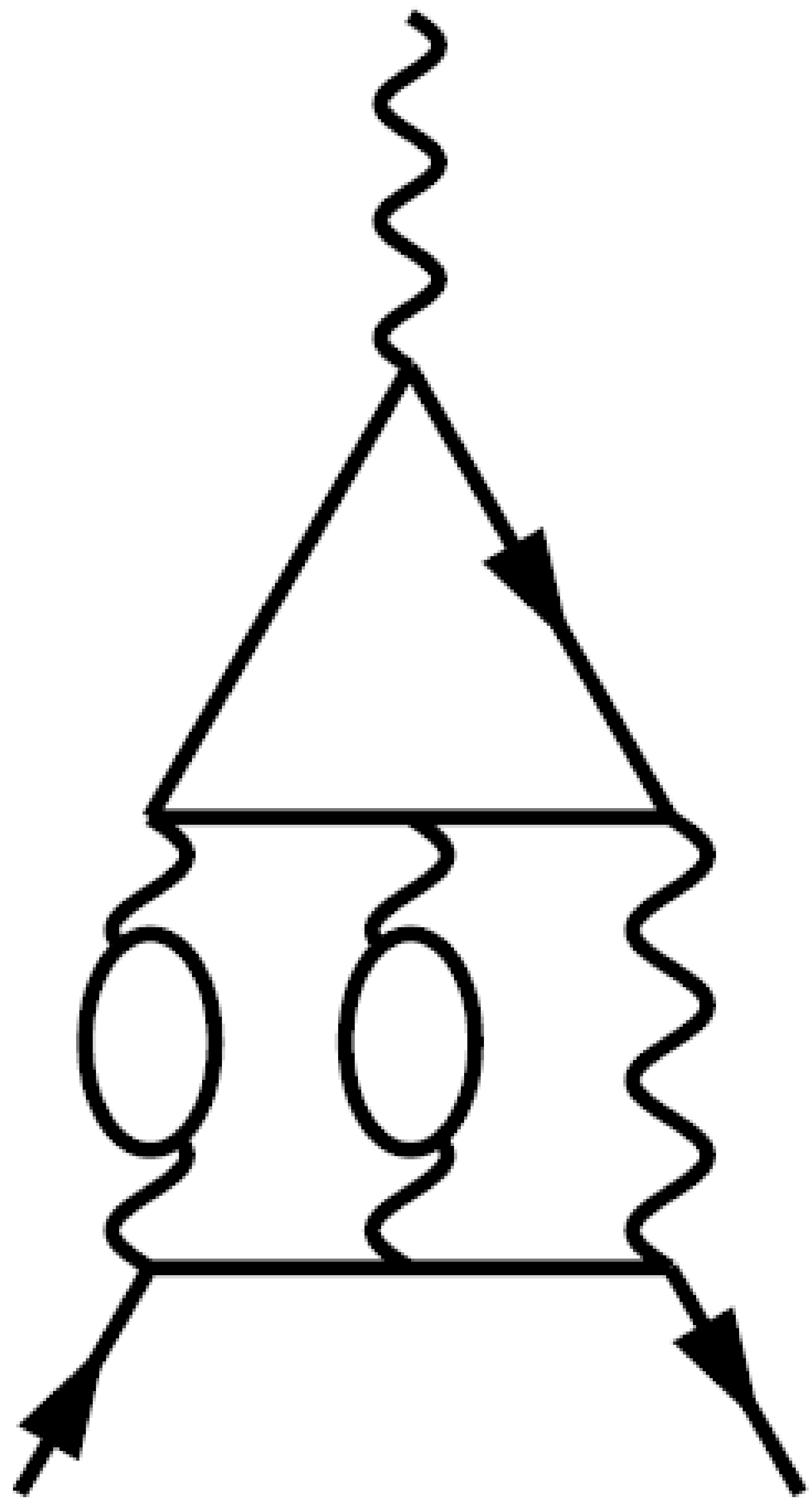}
\end{tabular}

~

~

\begin{tabular}{ccccccc}
Hadronic Vacuum Polarisation & Hadronic light-by-light & Weak \\
 (VP) & Scattering & Interactions \\
\includegraphics[width=0.2\linewidth]{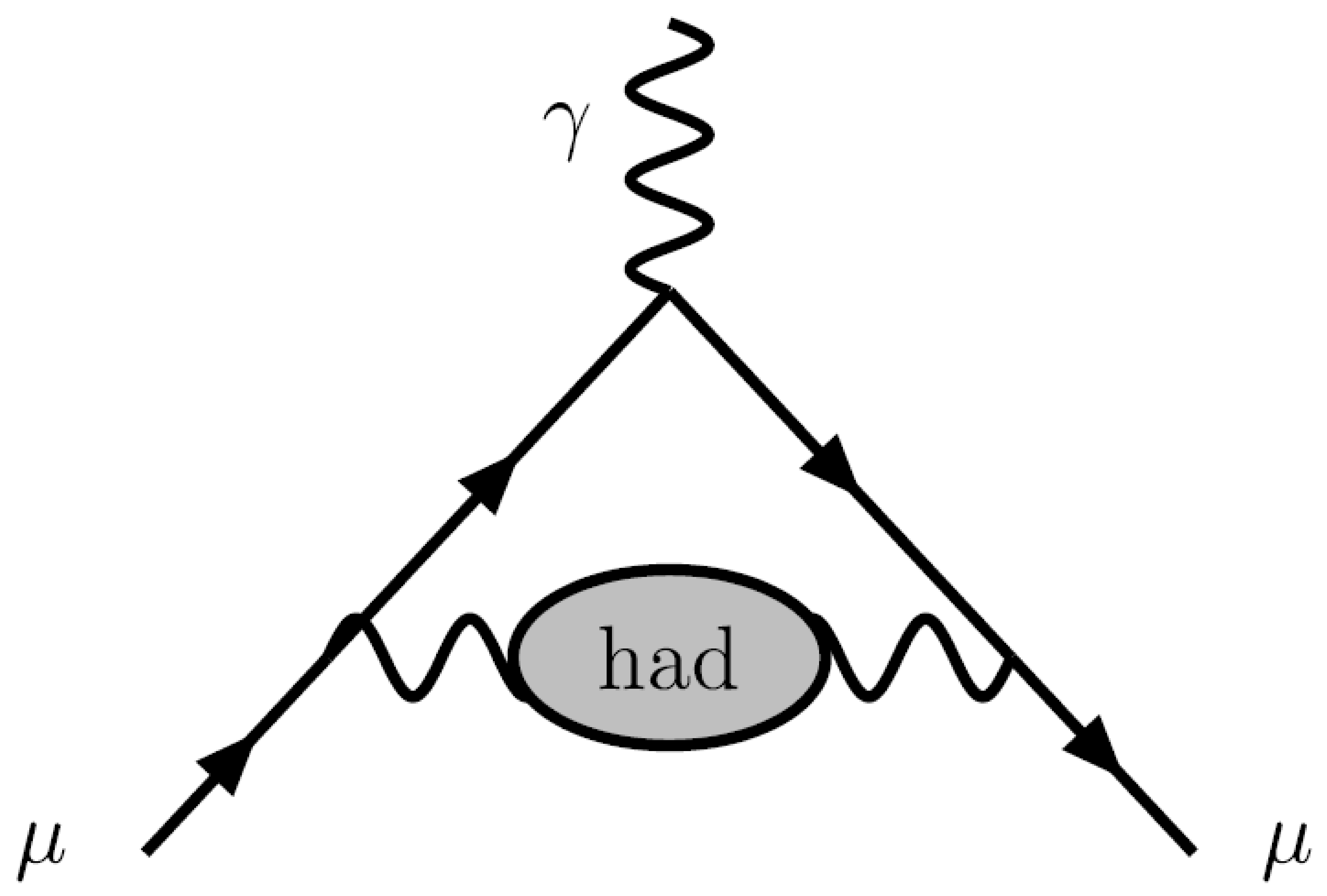}
&
\includegraphics[width=0.2\linewidth]{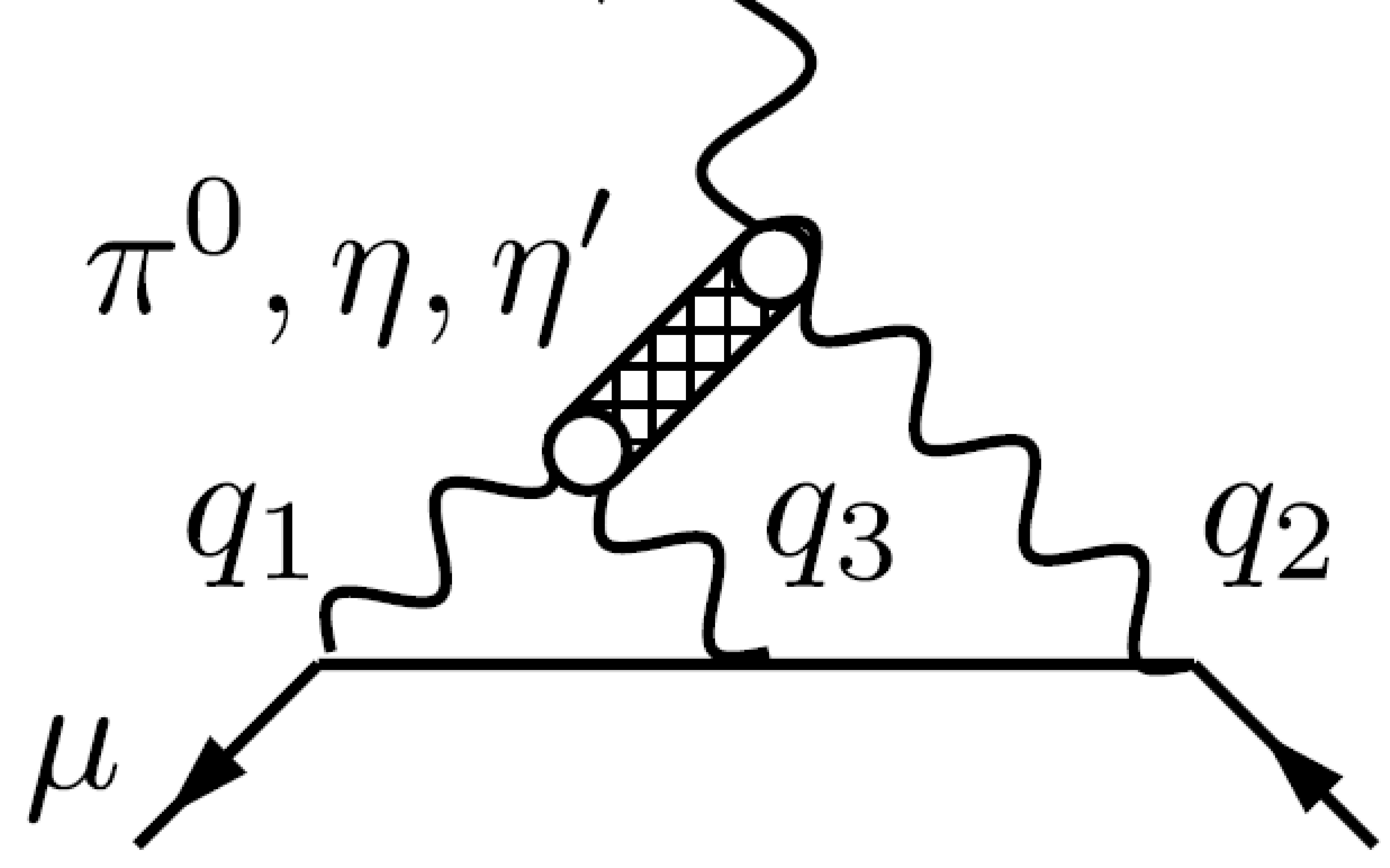}
&
\includegraphics[width=0.2\linewidth]{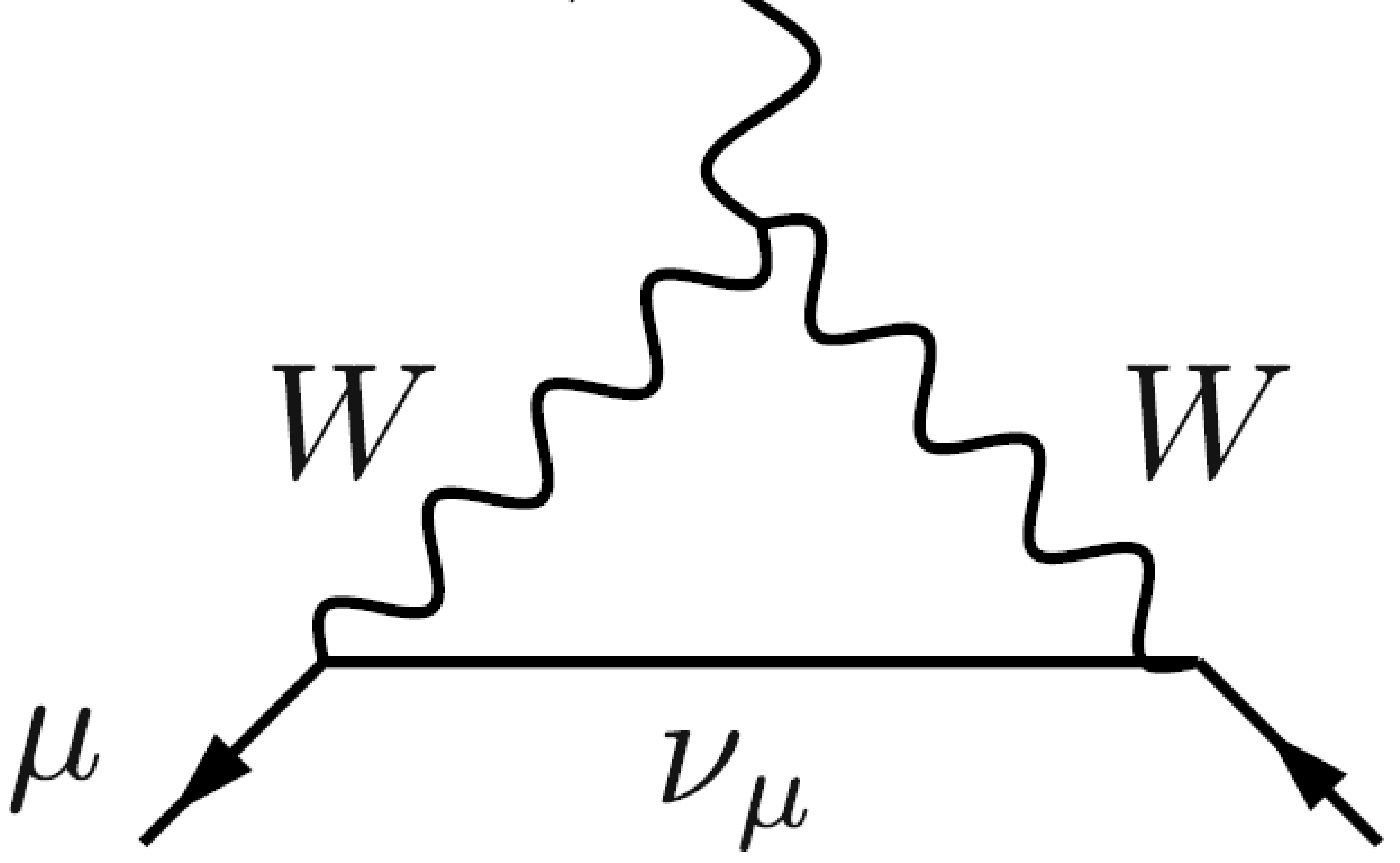}
\end{tabular}
\end{center}
\caption{Examples of diagrams contributing to the calculation of $a_\mu$.
Up: QED diagrams of various orders in $\alpha$.
Bottom: VP, LbL and weak-interaction contributions \cite{Jegerlehner:2009ry}.
\label{fig:a}
}
\end{figure}

Since the first measurement (for the electron) \cite{Nafe} and its
interpretation within the QED framework \cite{Schwinger}, both the
prediction and the measurement of $a$ have undergone a tremendous improvement
in precision, to the point that hadronic vacuum polarization (VP)
i.e. modifications of the photon propagator, hadronic light-by-light
scattering (LbL) and weak interactions must be taken into account
(Fig. \ref{fig:a}).
Understanding the value of $a_\mu$ necessitates a precise knowledge of
the value of the fine structure constant $\alpha$.
From the development \cite{Aoyama:2012wk} of $a_e$ and of $a_\mu$
\footnote{I have truncated the numerical factors.},
\begin{eqnarray*}
a_e &=& \gfrac{\alpha}{2\pi} 
-0.3 \left(\gfrac{\alpha}{\pi}\right)^2
+1.2 \left(\gfrac{\alpha}{\pi}\right)^3
-1.9 \left(\gfrac{\alpha}{\pi}\right)^4
+9.2 \left(\gfrac{\alpha}{\pi}\right)^5 + 1.7 \times 10^{-12} (\QCD + \weak),
\\
a_\mu &=& \gfrac{\alpha}{2\pi} 
+0.8 \left(\gfrac{\alpha}{\pi}\right)^2
+24. \left(\gfrac{\alpha}{\pi}\right)^3
+131. \left(\gfrac{\alpha}{\pi}\right)^4
+753. \left(\gfrac{\alpha}{\pi}\right)^5 +7.1 \times 10^{-8} (\QCD + \weak),
\end{eqnarray*}
%
%
we see that due to the $\mu$-to-$e$ mass difference, the development
for $a_e$ converges extremely rapidly and that the non-QED
contributions are very small: a precise value of $\alpha$ can be
extracted from $a_e$ and then injected in the calculation of $a_\mu$.
\begin{table}[Htb]
\begin{center} \small 
\begin{tabular}{cccccccc}
 $\alpha$ from & $a_\mu^{\QED}$ ($10^{-10}$) \\ 
\noalign{\vskip5pt}
\hline
 \noalign{\vskip5pt}
 $a_e$                    & 11 658 471.885 \pom 0.004 \\
\noalign{\vskip5pt}
Rubidium Rydberg constant & 11 658 471.895 \pom 0.008
\end{tabular}
\end{center}
\caption{Values of $a_\mu^{\QED}$ computed using values of $\alpha$ extracted from the measured value of $a_e$ and from atomic physics measurements \cite{Aoyama:2012wk}.
\label{tab:valeurs:amu}
}
\end{table}
The value of $a_\mu$ so obtained has a very small uncertainty and is
compatible with that obtained using a value of $\alpha$ from atomic physics 
(Table \ref{tab:valeurs:amu}):
the QED contribution, which has been computed up to the $5^{th}$ order in 
$\alpha$ \cite{Aoyama:2012wk}, is under excellent control.
Table \ref{tab:PDG:2014} presents the sizable contributions to the
prediction and the comparison with experiment as of 2014
\cite{PDG:2014}:
\begin{table}[Htb]
\begin{center} \small 
\begin{tabular}{lrl}
\hline
QED & 11 658 471.895 & \pom 0.008 \\
Leading hadronic vacuum polarization (VP) & 692.3~~~~ & \pom 4.2 \\
Sub-leading hadronic vacuum polarization & $-9.8$~~~~ & \pom 0.1 \\
Hadronic light-by-light (LbL) & 10.5~~~~ & \pom 2.6 \\
Weak (incl. 2-loops) & 15.4~~~~ & \pom 0.1 \\
\hline
Theory & 11 659 180.3~~~~ & \pom 4.2 \pom 2.6 \\
Experiment (E821 @ BNL) \cite{Bennett:2006fi} & 11 659 209.1~~~~ & \pom 5.4 \pom 3.3 \\
\hline
Exp. $-$ theory & +28.8~~~~ & \pom 8.0 \\
\end{tabular}
\end{center}
\caption{Contributions to the prediction for $a_\mu$ ($10^{-10}$) and
 comparison with experiment as of 2014
 \cite{PDG:2014}. \label{tab:PDG:2014}}
\end{table}

\begin{itemize}
\item
The QED contribution is the main contributor to the value of $a_\mu$,
while the uncertainty is dominated by the hadronic contributions (VP
and LbL);
\item
The uncertainties of the prediction and of the measurement are of similar magnitude;
\item
The measured value exceeds the prediction with, assuming Gaussian
statistics, a significance of $\approx 3.6$ standard deviations.
\end{itemize} 

As QCD is not suited to precise low energy calculations, the VP
contribution to $a_\mu$ is computed from the ``dispersion integral''
(\cite{Jegerlehner:2009ry} and references therein): 
\begin{eqnarray}
a_\mu^{\VP} = \left(\gfrac{\alpha m_\mu} {3\pi} \right)^2
\int{\gfrac{R(s) \times \hat{K}(s)}{s^2} \dd s}, 
\end{eqnarray}
where $ R(s) $ is the the cross section of $\epem$ to hadrons
at center-of-mass (CMS) energy squared $s$, normalized to the
pointlike muon pair cross section $\sigma_0$:
$ R(s) = \sigma_{\epem\to \hadrons} / \sigma_0$,
and $\hat{K}(s)$ is a known function that is of order unity on the $s$ range 
$[(2 m_\pi c^2)^2, \infty[$.
Technically, the low energy part of the integral is obtained from
experimental data (up to a value often chosen to be $E_{\cut} = 1.8 \gev$), while the
high-energy part is computed from perturbative QCD (pQCD).
Due to the presence of the $s^2$ factor at the denominator of the
integrand, the precision of the prediction of $a_\mu$ relies on
precise measurements at the lowest energies, and the channels
with the lightest final state particle rest masses,
 $\pip \pim$, 
 $\pip \pim \piz$, 
 $\pip \pim 2\piz$, $\pip \pim\pip \pim$, 
 $K K$ are of particular importance.

\section{\babar\ measurements: the ISR method}

\begin{figure}[Htb]
\includegraphics[width=0.7\linewidth, trim=0cm 9.7cm 0cm 0cm, clip]{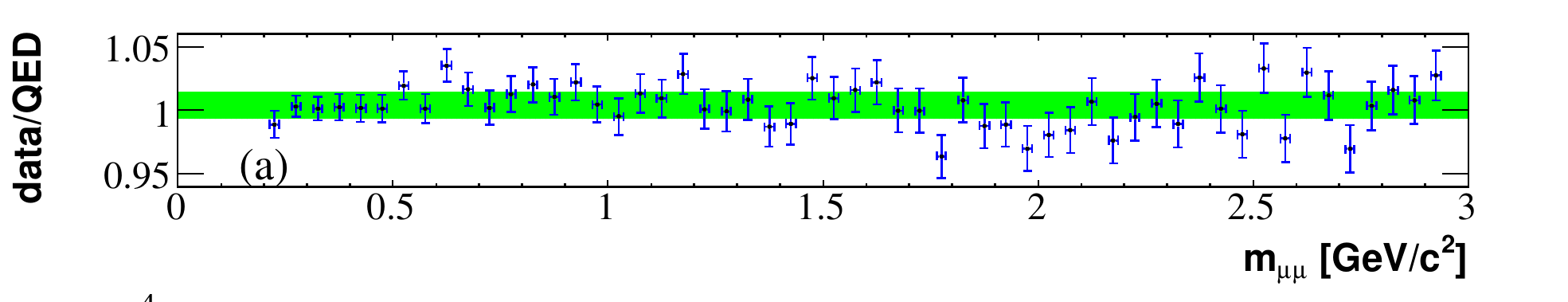}
\hfill
\includegraphics[width=0.29\linewidth]{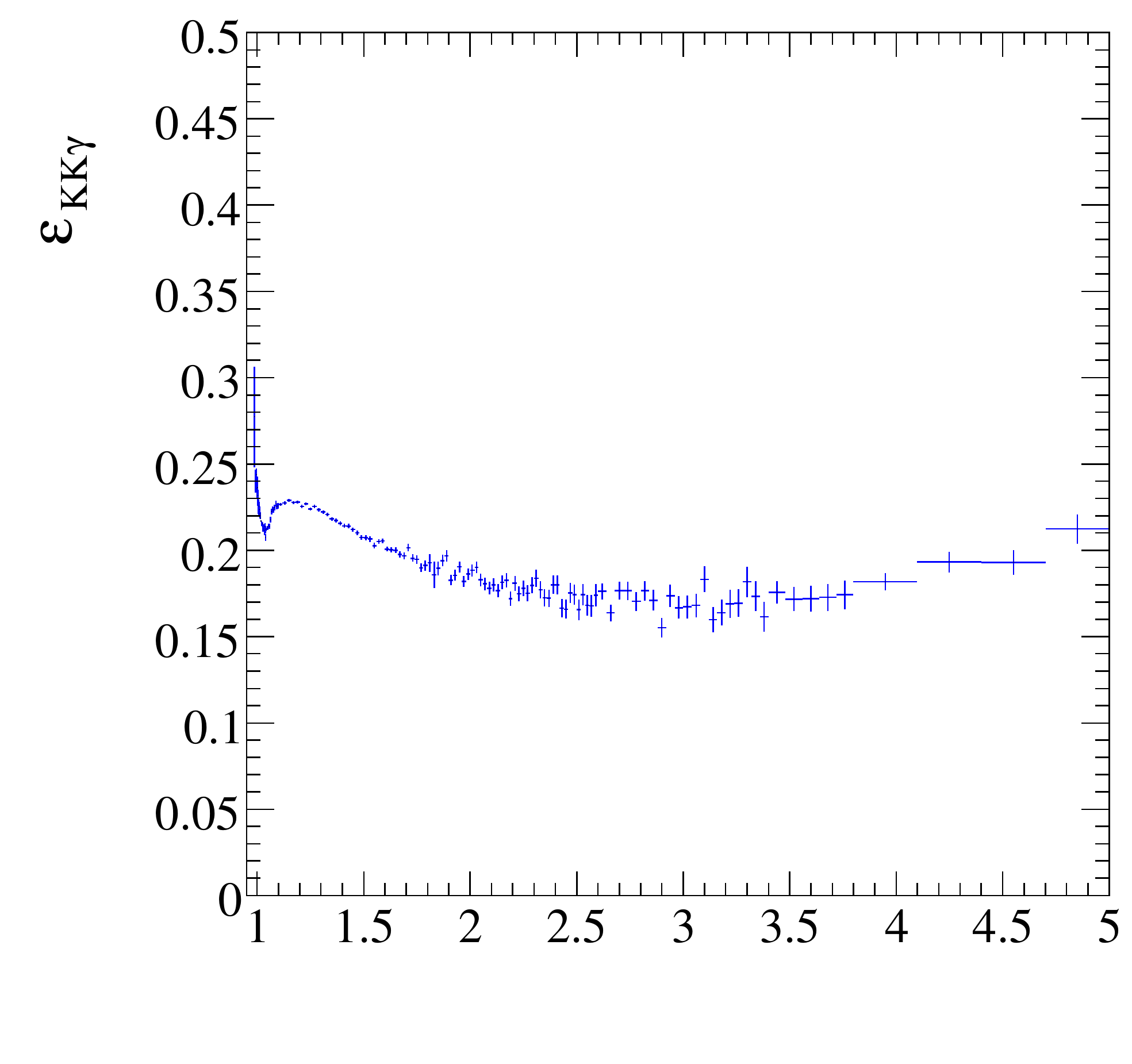}
\put(-40,0){\footnotesize $\sqrt{s'} (\gev)$}
\caption{Left: $\mu^+\mu^-$ cross section as a function of the $\mu^+\mu^-$ invariant mass compared to the QED prediction, as a sanity check for the \babar\ NLO analyses 
 \cite{Aubert:2009ad,Lees:2012cj,Lees:2013gzt}.
Right: The \babar\ acceptance for the $\Kp\Km$ analysis as a function of the $\Kp\Km$ invariant mass \cite{Lees:2013gzt}.
\label{fig:acceptanceKK}
}
\end{figure}

The \babar\ experiment \cite{Aubert:2001tu,TheBABAR:2013jta} at the
SLAC National Accelerator Laboratory has committed itself over the
last decade to the systematic measurement of the production of all
hadronic final states using the initial-state radiation (ISR) process.
The cross section of the $\epem$ production of a final state $f$ at a
CMS energy squared $s'$ can be obtained from the differential
cross section of the ISR production $\epem \to f ~ \gamma$ through the
expression:
\begin{eqnarray}
\gfrac{\dd \sigma_{[\epem \to f ~ \g]}}{\dd s'} (s')
=
\gfrac{2m}{s} W(s, x) \sigma_{[\epem \to f]} (s'),
\end{eqnarray}
where $W(s, x)$, the probability  density  to radiate a photon with
energy $E_\g = x\sqrt{s}$, is a known ``radiator'' function
\cite{Bonneau:1971mk}, and $\sqrt{s}$ is here the CMS energy of the
initial \epem pair, which is close to 10.6\,GeV for \babar.
In contrast with the energy scans that provided the earlier
experimental information on the variations of $R$
(see Figs. 50.5 and 50.6 in Ref. \cite{PDG:2014} and references in
their captions),
this ISR method makes an optimal use of the available luminosity and
allows a consistent measurement over the full energy range with the same
accelerator and detector conditions.
In addition, in the case of \babar\, the \epem initial state is
strongly boosted longitudinally so the detector acceptance stays
sizable down to threshold (Fig. \ref{fig:acceptanceKK} right).

The observation of the hadronic final state alone, if kinematically compatible with
a system recoiling against a single massless particle, would allow the
reconstruction of the event and the measurement of $s'$, but when in
addition the ISR photon is observed ($\gamma$-tagging), a powerful
background rejection and a good signal purity can be achieved.
We have performed most of these measurements using a leading-order
(LO) method, in which the final state $f$ and the ISR photon are
reconstructed regardless of the eventual presence of additional
photons.
For these analyses the differential luminosity is obtained from the
luminosity of the collider, known with a typical precision of $1 \%$,
and involves a computation of the detection efficiency that relies on
Monte Carlo (MC) simulations\footnote{
A review on the {\tt PHOKHARA} and {\tt AfkQed} event generators used
in our {\tt GEANT4}-based simulations can be found in section 21 of
Ref. \cite{Bevan:2014iga}.}
 \cite{Aubert:2004kj,Aubert:2006jq,Aubert:2007uf,Aubert:2007ef,Aubert:2007ym},
 \cite{Lees:2012cr,Lees:2011zi,Lees:2013uta,Druzhinin:2007cs,Lees:2014xsh,Lees:2013ebn,Lees:2015iba}.
This experimental campaign has lead \babar\ to improve the precision
of the contribution to $a_\mu$ of most of the relevant channels by a
large factor, typically close to a factor of three.

A list of the contributions $a_\mu^f$ to $a_\mu^{\VP}$ for a number of
individual hadronic final states $f$, available at the time, can be
found in Table 2 of Ref. \cite{Davier:2010nc}.

\section{BaBar NLO ($\epem \to f ~ \gamma ~ (\gamma))$ results}

\begin{figure}[Htb]
\hfill
\includegraphics[width=0.45\linewidth]{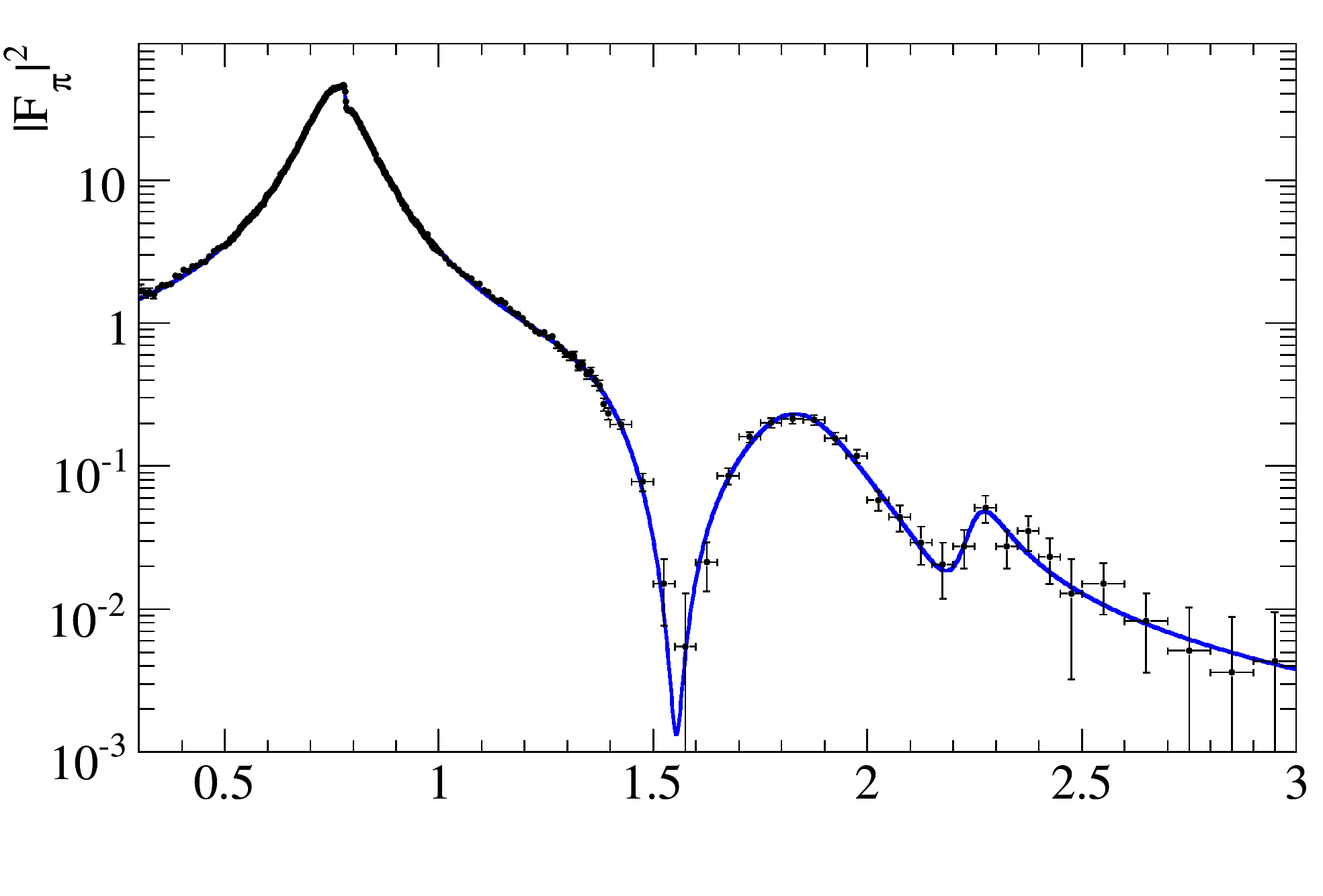}
\hfill
\includegraphics[width=0.45\linewidth]{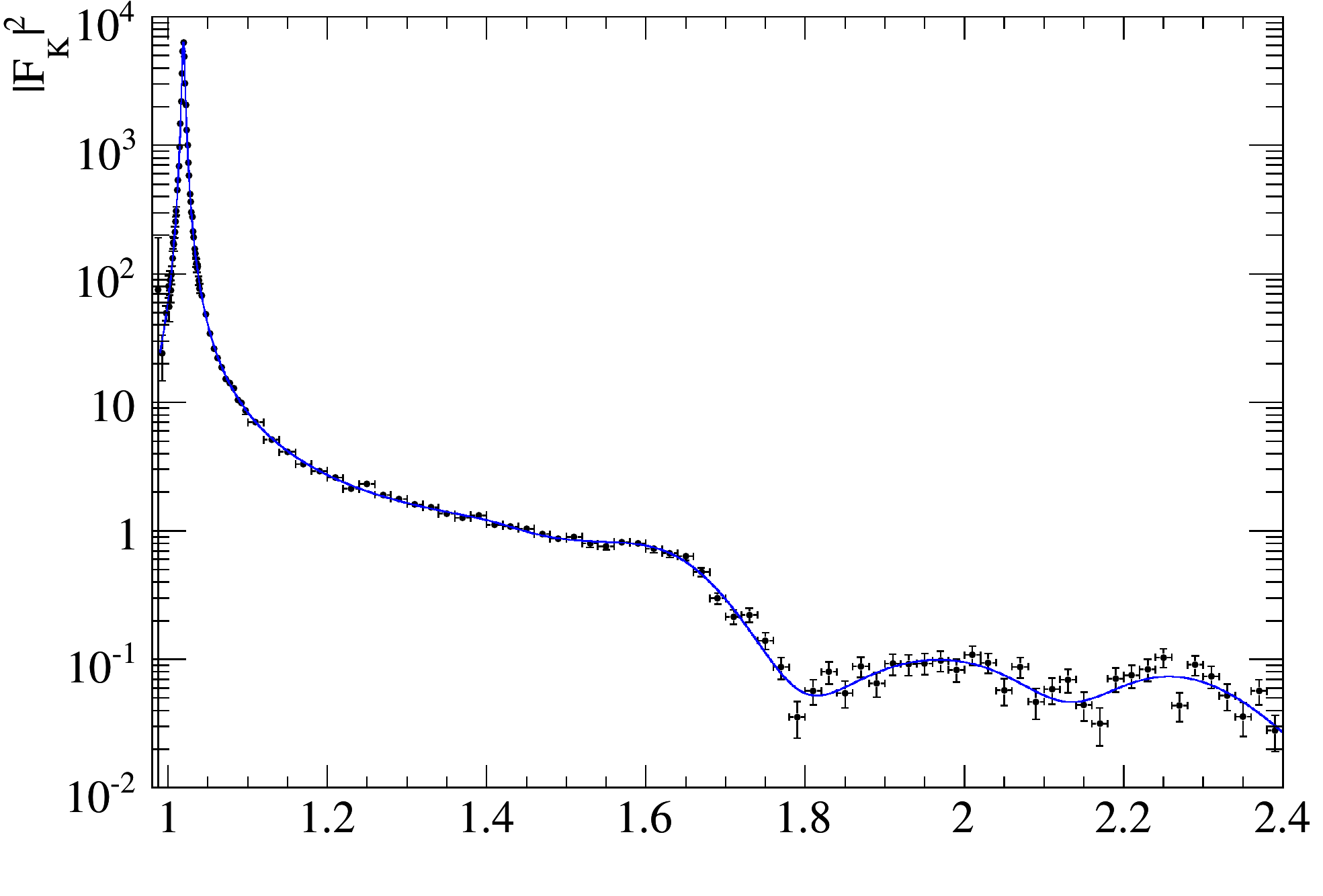} 
\hfill
~ 
\put(-300,90){$\pip\pim$}
\put(-70,90){$\Kp\Km$}
\put(-60,-5){\footnotesize $\sqrt{s'} (\gev)$}
\put(-270,-2){\footnotesize $\sqrt{s'} (\gev)$}
\caption{\babar\ NLO measurements:
Vector dominance model (VDM) fits of the squared form-factors using a Gounaris-Sakurai (GS) parametrization.
Left:$\pip\pim$ \cite{Aubert:2009ad,Lees:2012cj}.
Right: $\Kp\Km$ \cite{Lees:2013gzt}.
\label{fig:VDM}
}
\end{figure}

\babar\ has also developed a new method that was applied to
the dominant channel $\pip\pim$ \cite{Aubert:2009ad,Lees:2012cj}
and more recently to the $\Kp\Km$ channel \cite{Lees:2013gzt}.
The control of the systematics below the \% level made it necessary to
perform the analysis at the NLO level, that is, to take into account
the possible radiation of an additional photon, be it from the initial (ISR) or from the final (FSR) state.
The impossibility to control the global differential luminosity with
the desired precision, in particular the MC-based efficiency, lead us
to derive the value of $R$ from the ratio of the ISR production of the
final state $f$ to the ISR production of a pair of muons,
$\mu^+\mu^-$.
Most of the systematics, including those related to the absolute
luminosity, of the ISR photon reconstruction and of additional ISR
radiation, cancel in the ratio.
Figure \ref{fig:VDM} shows the obtained form-factor
(here squared) distributions extracted from the cross-section distributions,
together with fits
using the GS parametrization of the VDM model.
The values of $a_\mu^{\pip\pim}$ and of $a_\mu^{\Kp\Km}$ integrated over the most
critical range, that is, from threshold to 1.8\,GeV are more precise than the
average of the previous measurements (Table \ref{tab:comparison}).
\begin{table} [Htb] \footnotesize
\begin{tabular}{l|lllllllll} 
\hline
 & $\pip\pim$ & $\pip\pim\pip\pim$ & $\Kp\Km$ 
\\
\hline
\babar\ & $514.1 \pm 2.2 \pm 3.1$ \cite{Aubert:2009ad,Lees:2012cj} & $22.93 \pm 0.18 \pm 0.22 \pm 0.03 $ \cite{Lees:2012cr} & $13.64 \pm 0.03 \pm 0.36$ \cite{Lees:2013gzt} 
\\
Previous average
\cite{Davier:2010nc} & $503.5 \pm 4.5$ & $21.63 \pm 0.27 \pm 0.68$ & $13.35 \pm 0.10 \pm 0.43 \pm 0.29$
\\
Their difference $\Delta$ & $\! +10.6 \pm 5.9$ & $\!+ 1.30 \pm 0.79$ & $\! +0.29 \pm 0.63$
\end{tabular}
\caption{Contributions to $a_\mu$ for recent \babar\ publications: comparison of the measured value to the previous world average on the energy range $\sqrt{s'}< 1.8\,\gev$ (units $10^{-10}$).
\label{tab:comparison}
}
\end{table}

Even though
 neither the time-integrated luminosity nor the absolute
acceptance/efficiency were used in these precise $\pip\pim$ and
$\Kp\Km$ cross-section measurements, we checked that we understand
them by comparing the $\mu^+\mu^-$ cross section distribution we
observe to the QED prediction: a good agreement is found
(Fig. \ref{fig:acceptanceKK} left) within $0.4 \pom 1.1 \%$, which
is dominated by the large uncertainty on the time-integrated
luminosity ($\pom 0.9 \%$).

These NLO analyses were performed assuming that the FSR corrections
for the hadronic channel are negligible, as theoretical estimates are
well below the  systematic uncertainties in the cross section
\cite{Aubert:2009ad,Lees:2012cj,Lees:2013gzt}.
We have validated this assumption by an experimental study of the
ISR-FSR interference in $\mu^+\mu^-$ and $\pip\pim$ ISR production.
Because charge parities of the final state pair are opposite for ISR
and FSR, the interference between ISR and FSR changes sign with the
charge interchange of the two muons (pions).
As a consequence, investigation of the charge asymmetry of the process
gives access to the interference between ISR and FSR, which
enables the separate measurement of the magnitudes of the ISR and of
the FSR amplitudes \cite{Lees:2015qna}.
For the pion channel, results match a model where final state radiation
originates predominantly from the quarks that subsequently hadronize
into a pion pair, while 
for the muon control channel, good consistency is found with QED.

\section{Recent BaBar LO ($\epem \to f ~ \gamma)$ results}

\begin{figure}[Htb]
\includegraphics[width=0.24\linewidth]{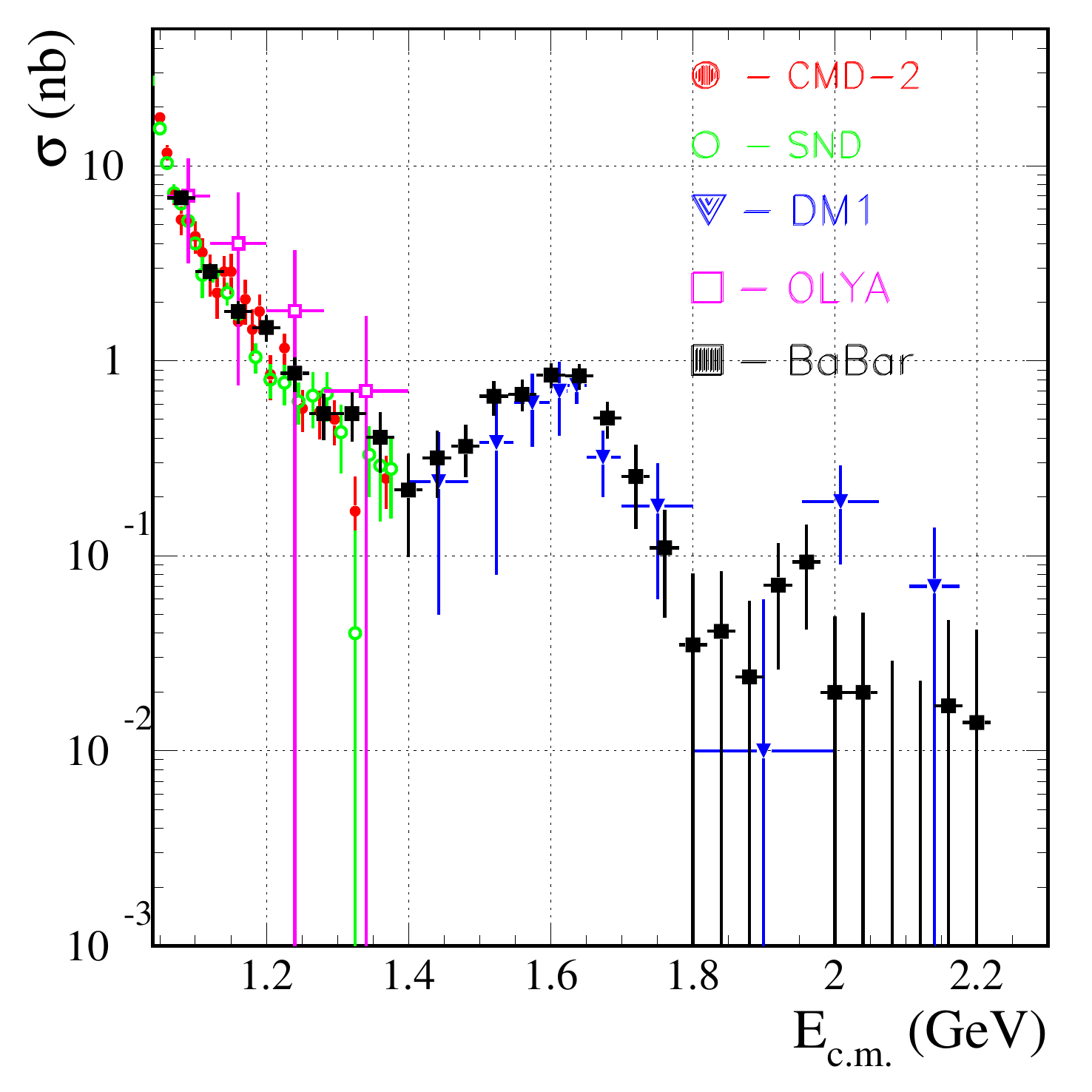}
\includegraphics[width=0.24\linewidth]{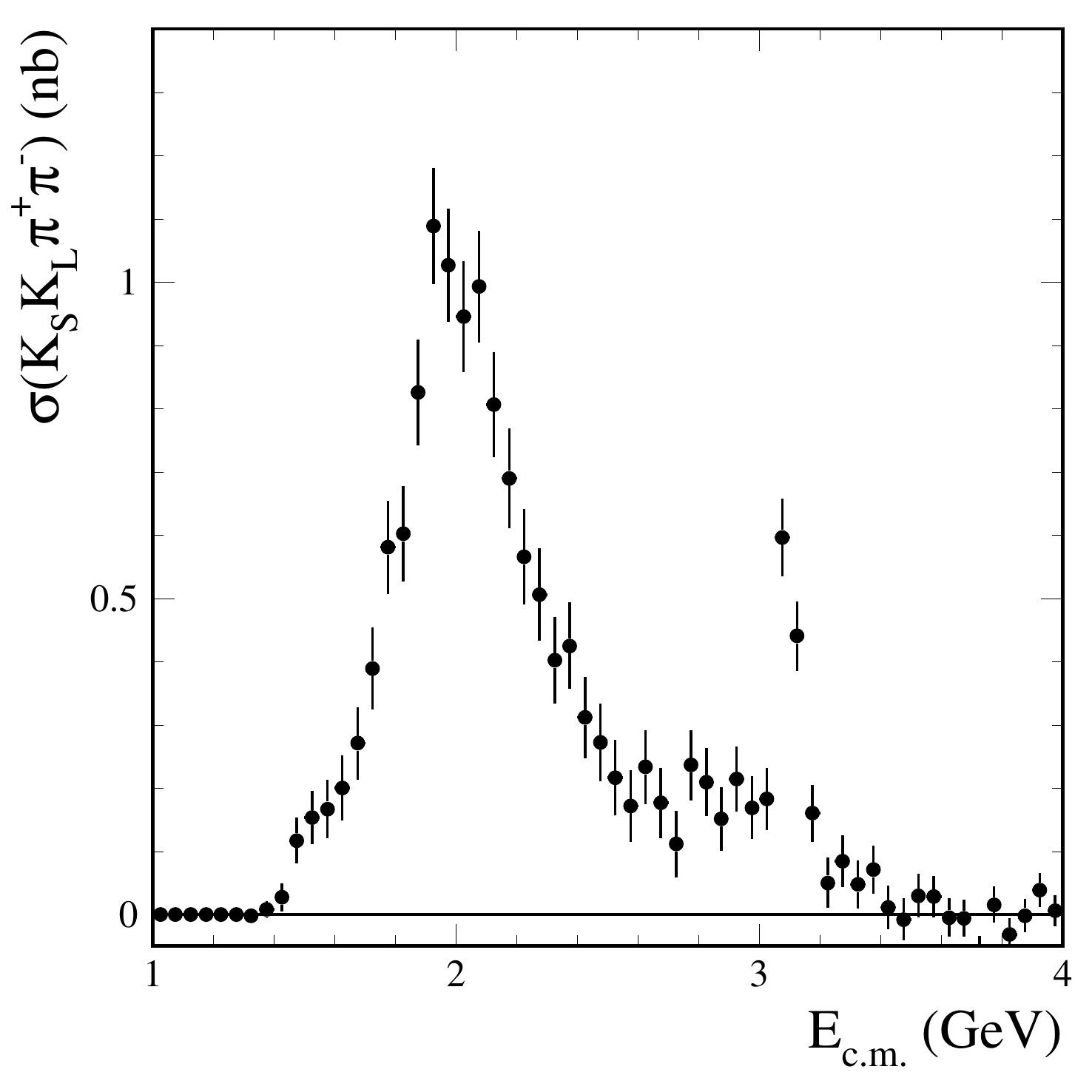}
\includegraphics[width=0.24\linewidth]{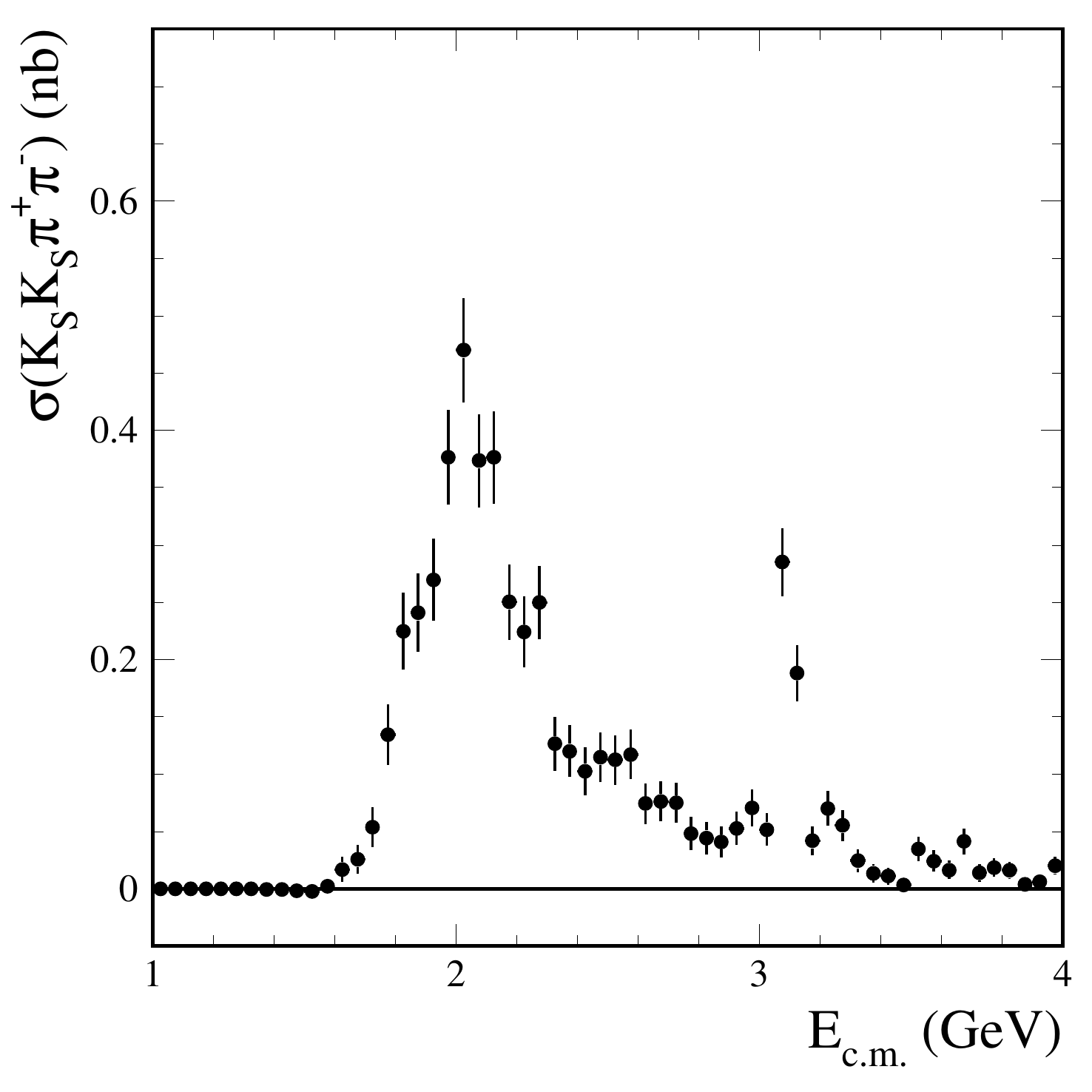} 
\includegraphics[width=0.24\linewidth]{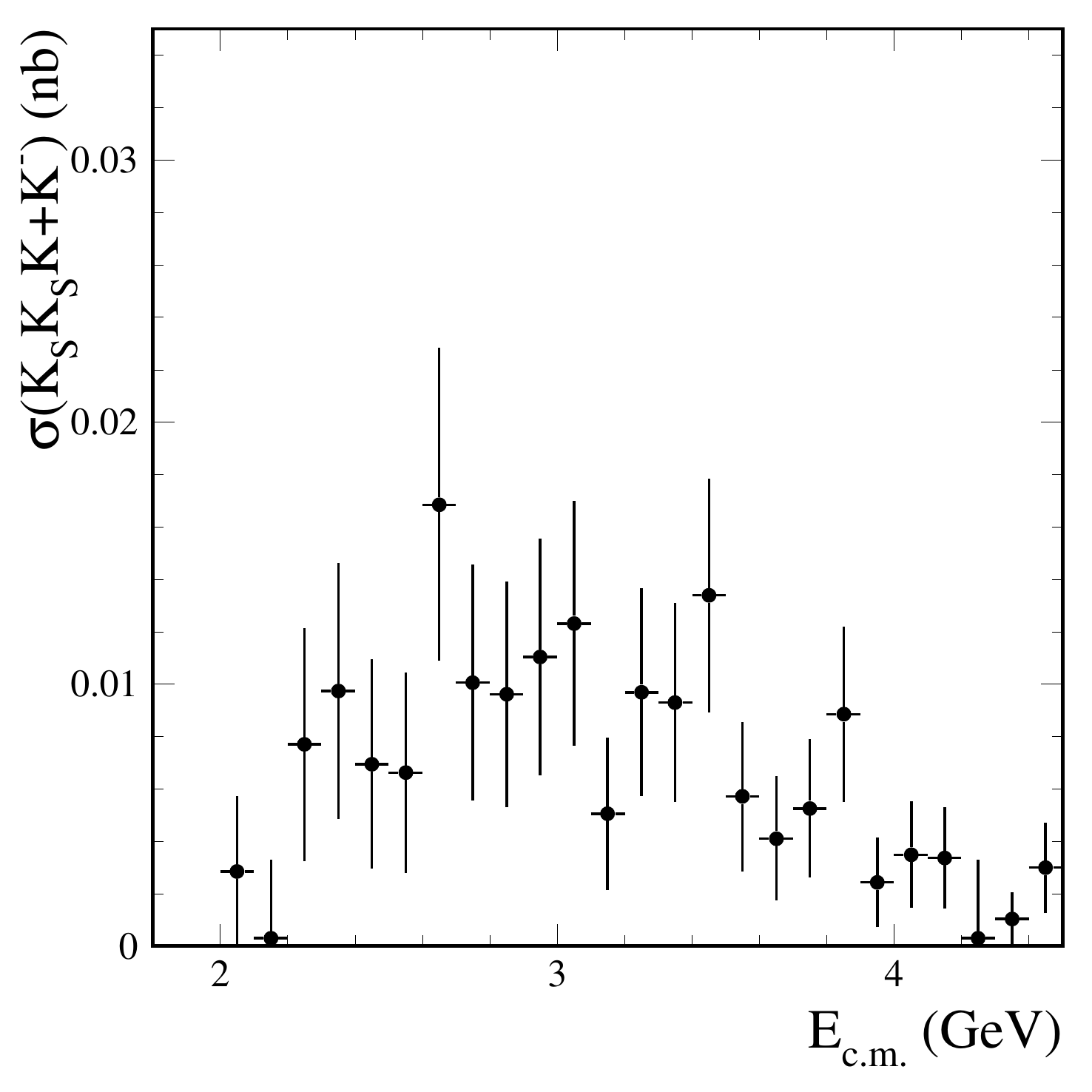}
 \put(-440,90){$\KS\KL$}
 \put(-300,90){\Magenta{$\KS\KL \pip\pim$}}
 \put(-185,90){\Magenta{$\KS\KS \pip\pim$}}
 \put( -70,90){\Magenta{$\KS\KS \Kp\Km$}}

\medskip \Black{~}

\hfill
\includegraphics[width=0.24\linewidth]{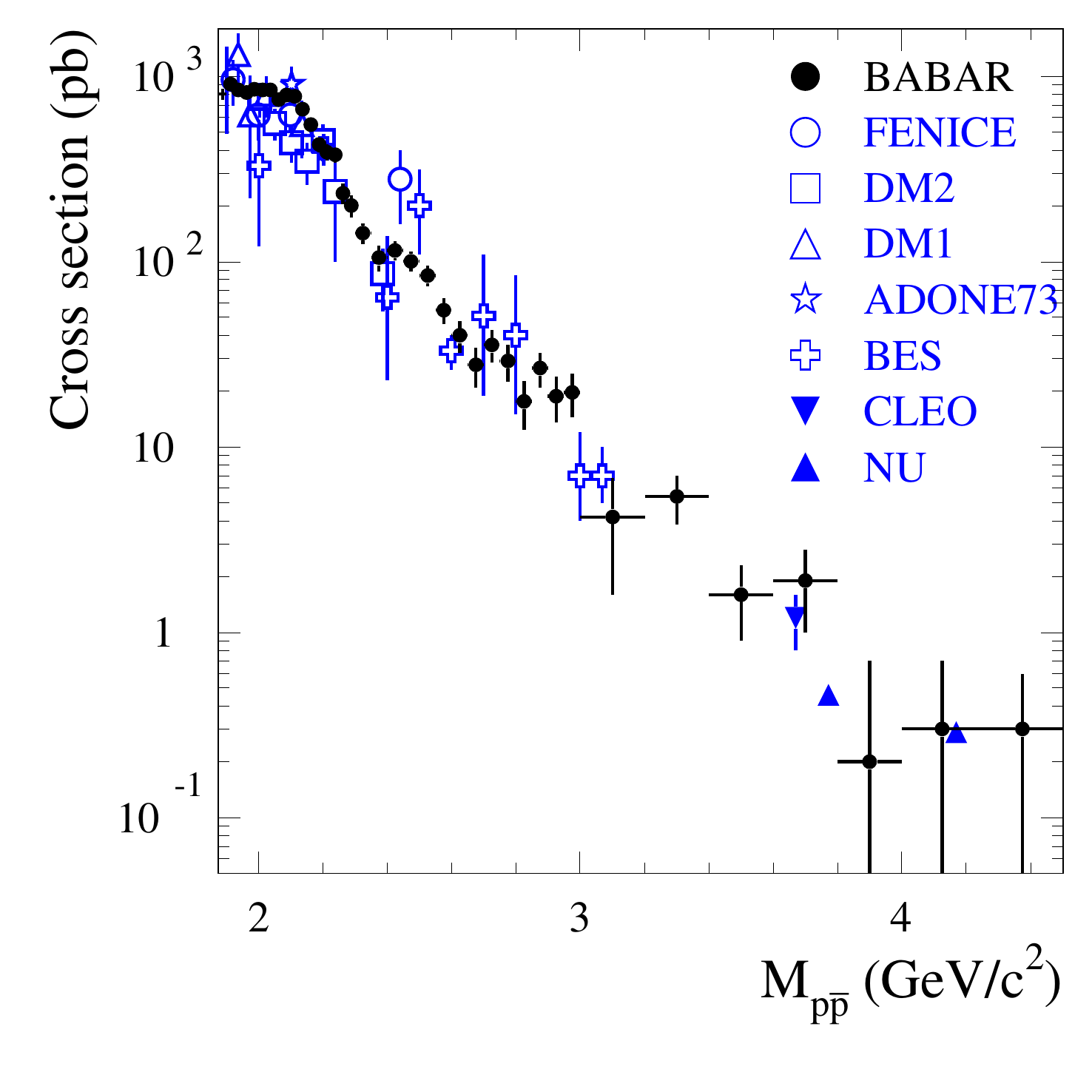}
\hfill
\includegraphics[width=0.24\linewidth]{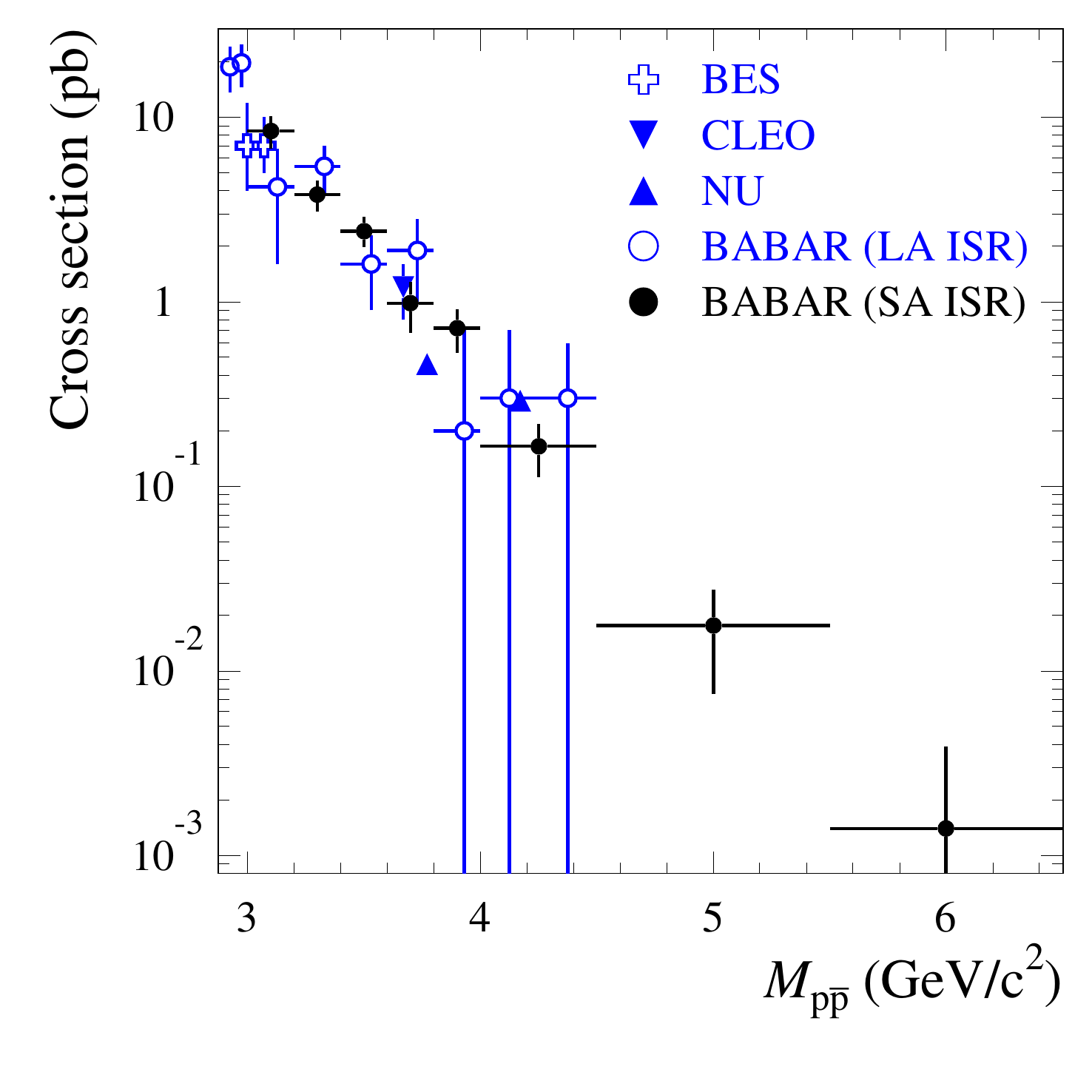}
\hfill
\includegraphics[width=0.24\linewidth]{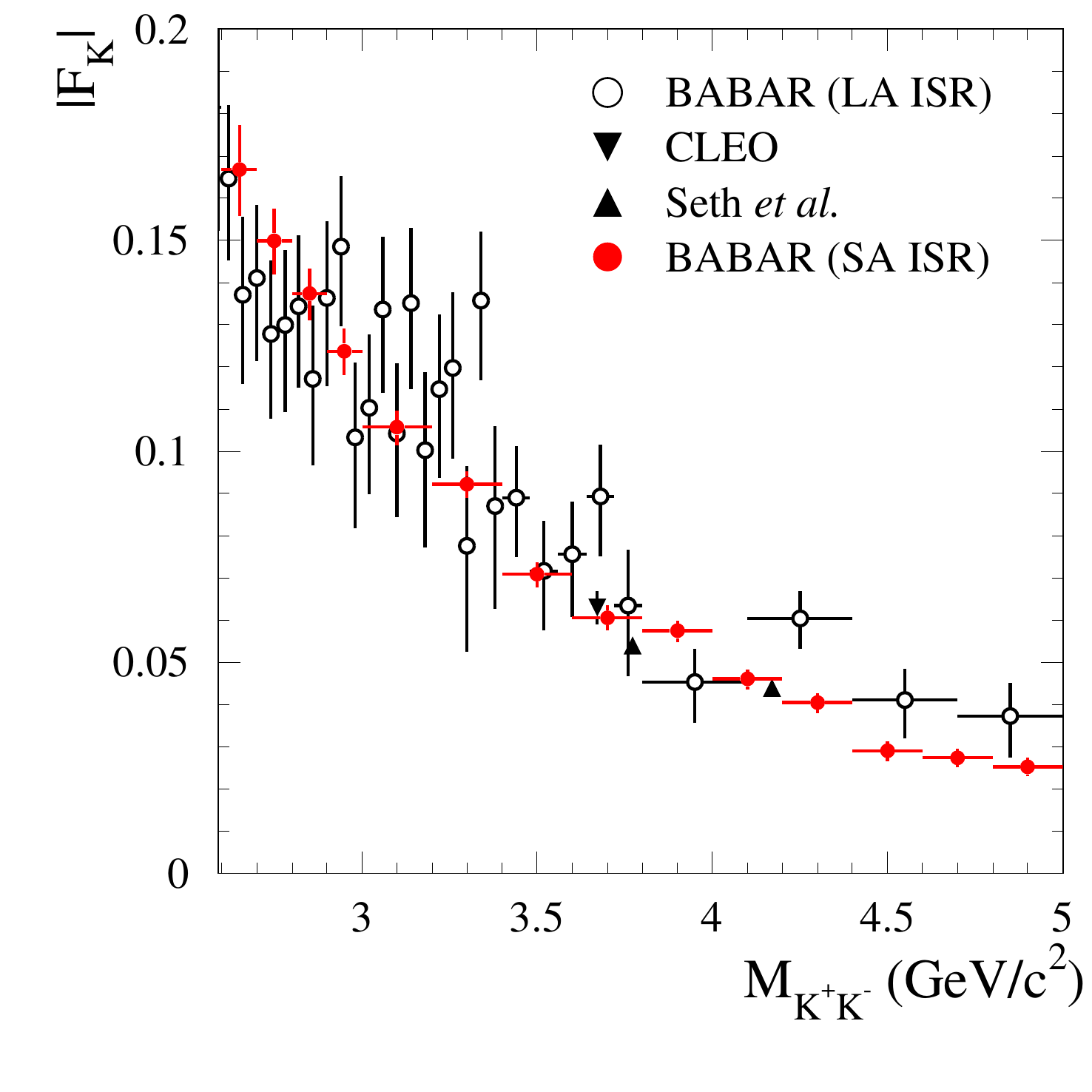}
\hfill
~
 \put(-410,40){$p \antiproton$}
 \put(-210,65){$p \antiproton$}
 \put( -75,65){$\Kp\Km$}

\hfill
\includegraphics[width=0.24\linewidth]{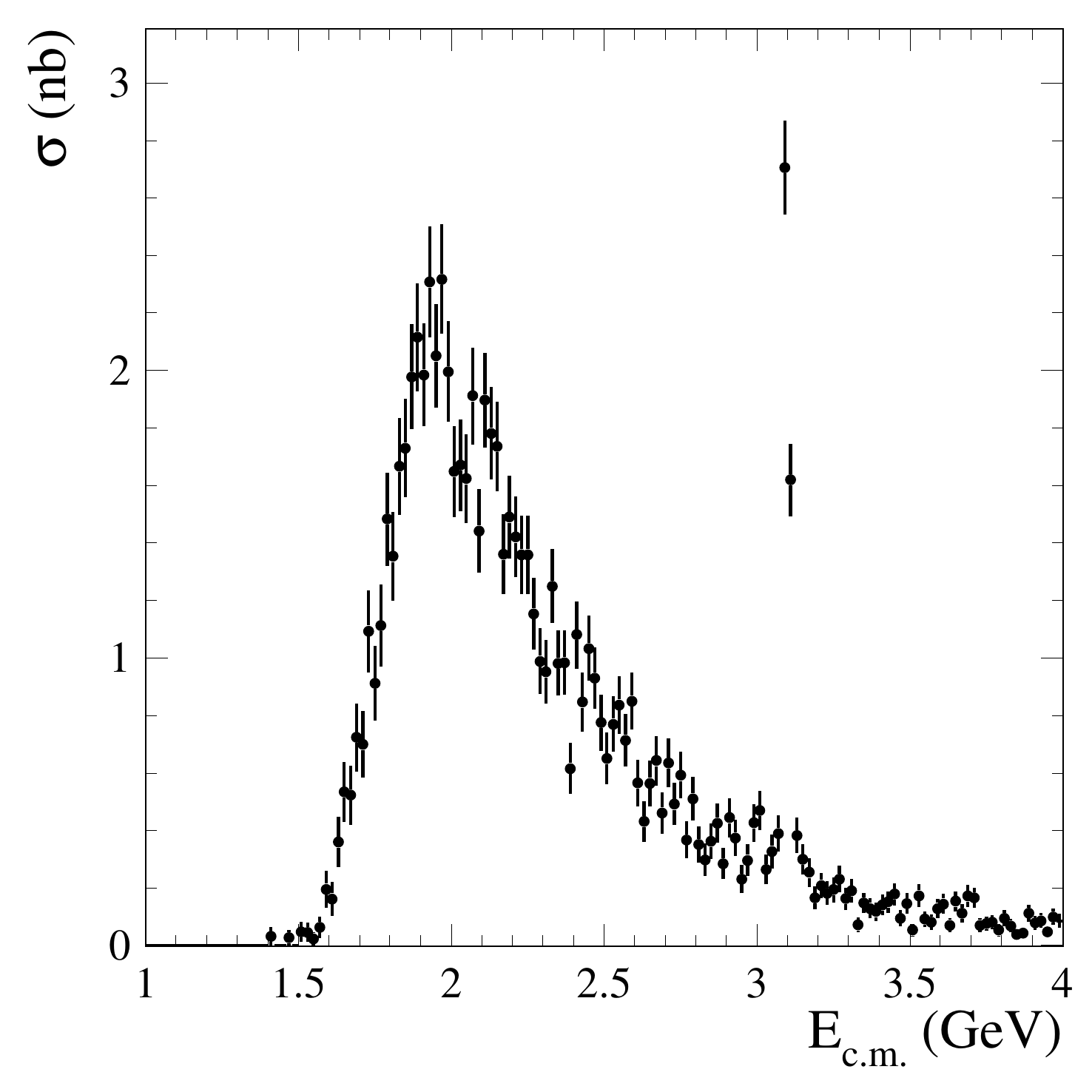}
\hfill
\includegraphics[width=0.24\linewidth]{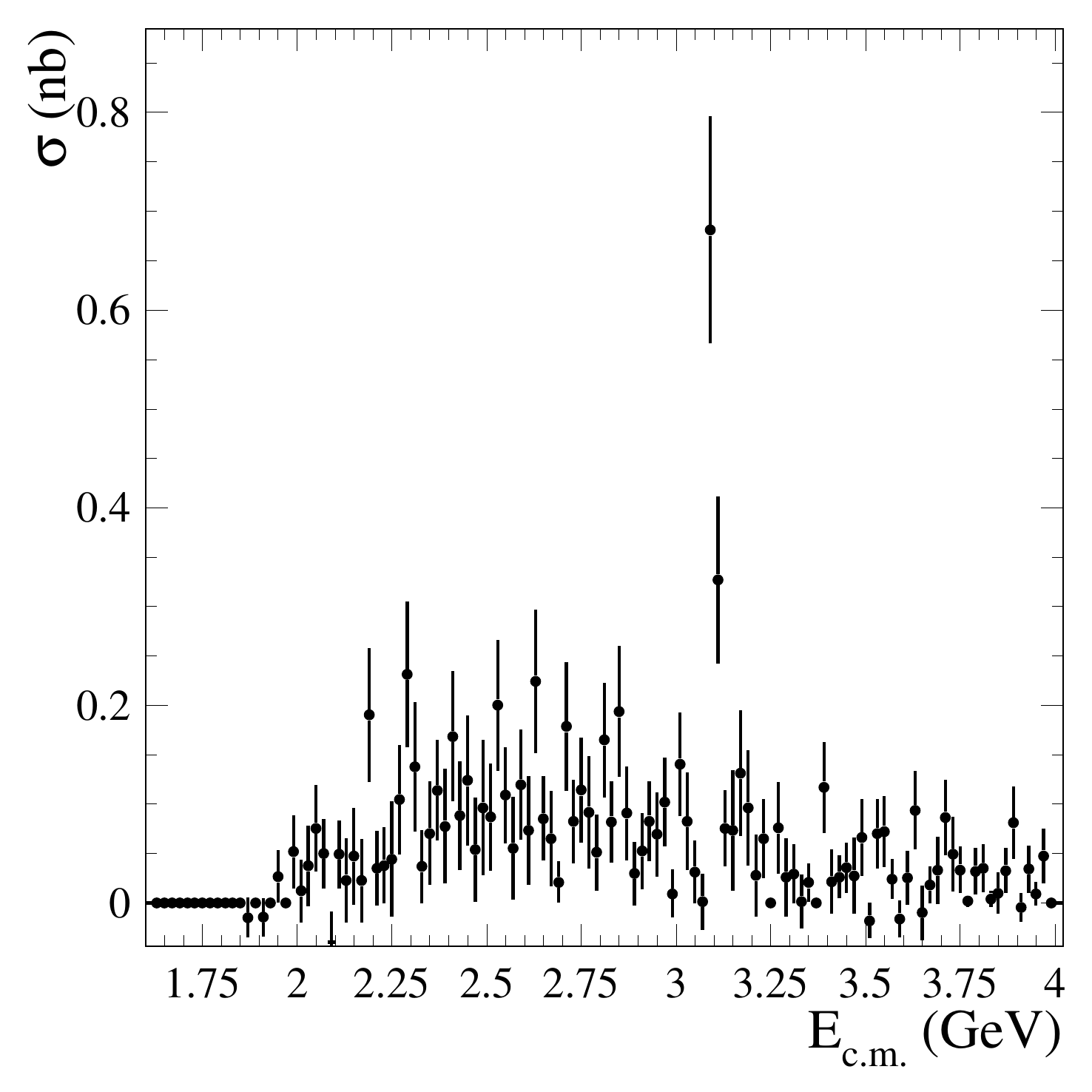}
\hfill
~
 \put(-350,80){\Magenta{$\KS\Kp\pim \piz$}}
 \put(-167,70){\Magenta{$\KS\Kp\pim \eta$}}
 \put(-340,65){\Red{preliminary}}
 \put(-155,55){\Red{preliminary}}
 \Black{~}
\caption{Recent LO results.
\Magenta{Magenta: First measurements}.
\Black{~}
Up: channels with two neutral kaons \cite{Lees:2014xsh}.
Center: $p \antiproton$ with \cite{Lees:2013ebn} and without \cite{Lees:2013uta} $\gamma$ tagging, and $\Kp\Km$ without \cite{Lees:2015iba} $\gamma$ tagging.
Bottom: $\KS\Kp\pim h^0$, the neutral meson $h^0$ being either a $\piz$ or an $\eta$ (preliminary).
\label{fig:recent:LO}
}
\end{figure}

\begin{figure}[Htb]
\hfill
 \includegraphics[width=0.42\textwidth]{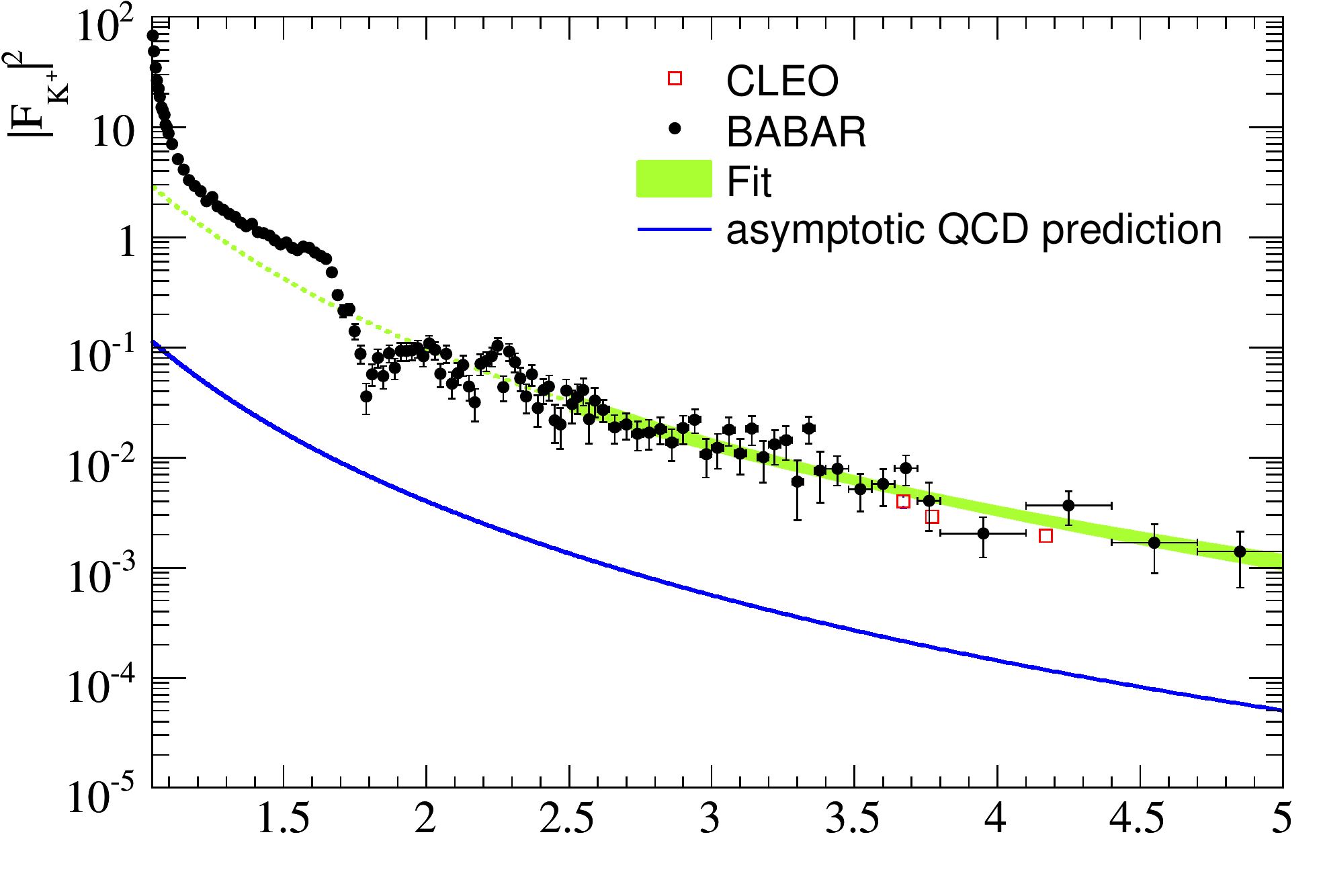}
\hfill
 \includegraphics[width=0.34\textwidth]{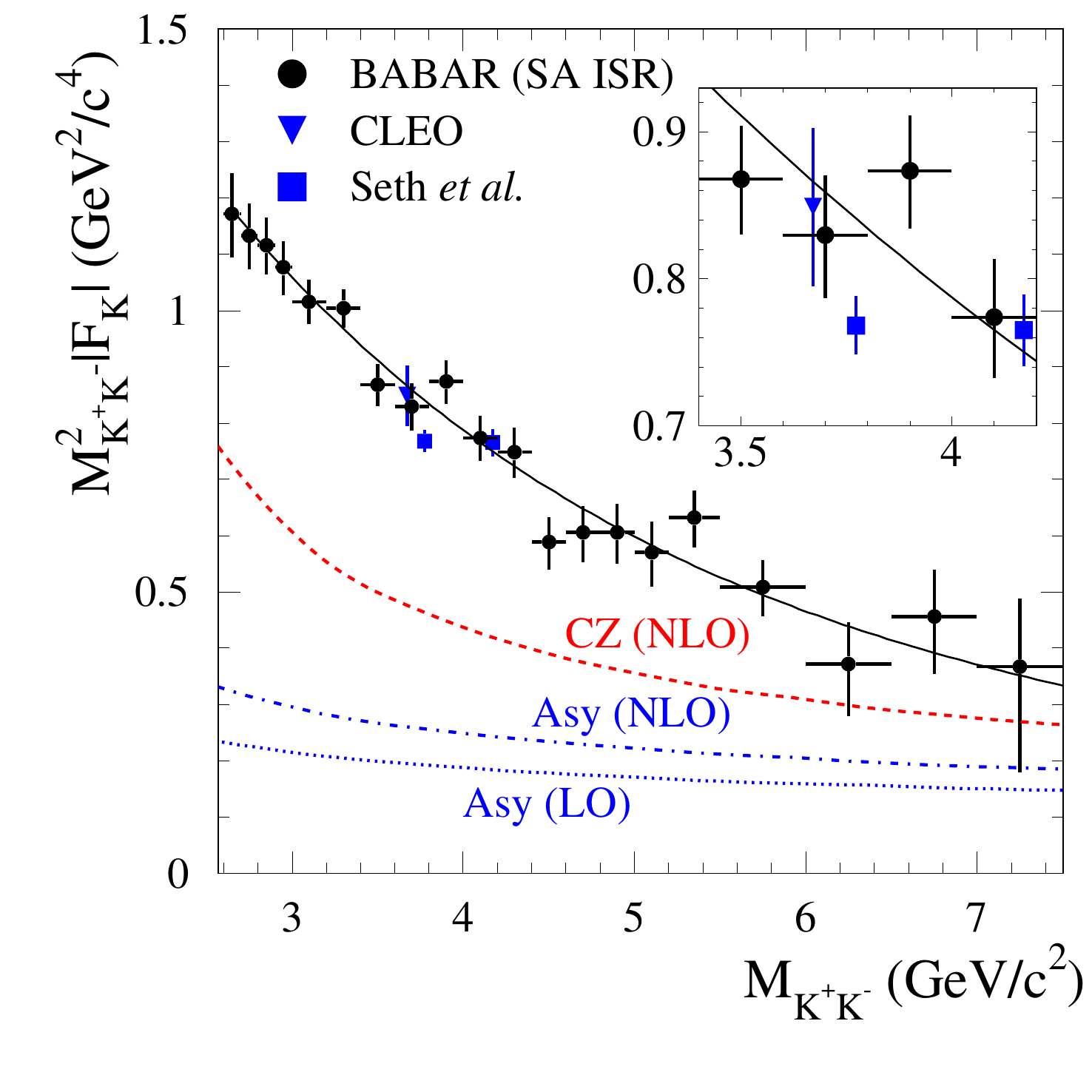} 
\hfill
~
 \put(-410,40){\Blue \scriptsize CZ (NLO) \Black}
\caption{Comparison of the \babar\ $\Kp\Km$ results with 
 Chernyak-Zhitnitsky \cite{Chernyak:1977fk} pQCD predictions.
With (left, \cite{Lees:2013gzt}) and without (right, \cite{Lees:2015iba}) $\gamma$ tagging.
\label{fig:KK:pQCD}
}
\end{figure}

Recently \babar\  obtained results on channels with two neutral kaons
$\KS\KL$, 
$\KS\KL \pip\pim$, 
$\KS\KS \pip\pim$ and 
$\KS\KS \Kp\Km $ \cite{Lees:2014xsh}
(Fig. \ref{fig:recent:LO} up),
 on
$\KS\Kp\pim \piz$ and $\KS\Kp\pim \eta$ (preliminary)
(Fig. \ref{fig:recent:LO} bottom),
and updated the $p \bar{p}$ analysis to the full statistics \cite{Lees:2013ebn}
(Fig. \ref{fig:recent:LO} center left).
The $p \bar{p}$ measurement has also been extended up to 6.5 \gev \cite{Lees:2013uta}
(Fig. \ref{fig:recent:LO} center center)
and the $\Kp\Km$ measurement to 8 \gev \cite{Lees:2015iba}
(Fig. \ref{fig:recent:LO} center right)
by untagged analyses.

pQCD is found to fail to describe the $\Kp\Km$ form factors
extracted from our cross section measurements
(Fig. \ref{fig:KK:pQCD}), but there is some hint that the discrepancy
is getting better at higher mass, which kind-of supports the use of
pQCD for the calculation of the dispersion integral above $E_{\cut}$.
Note that given the improvement in precision of the hadronic cross
sections, the most recent prediction 
\cite{Jegerlehner:2015stw}
restricts the $s$ range over which
pQCD is used to [4.5 -- 9.3]\,GeV and [13\,GeV -- $\infty$[.

\begin{figure}[Htb]
\centerline{
\includegraphics[width=0.85\linewidth]{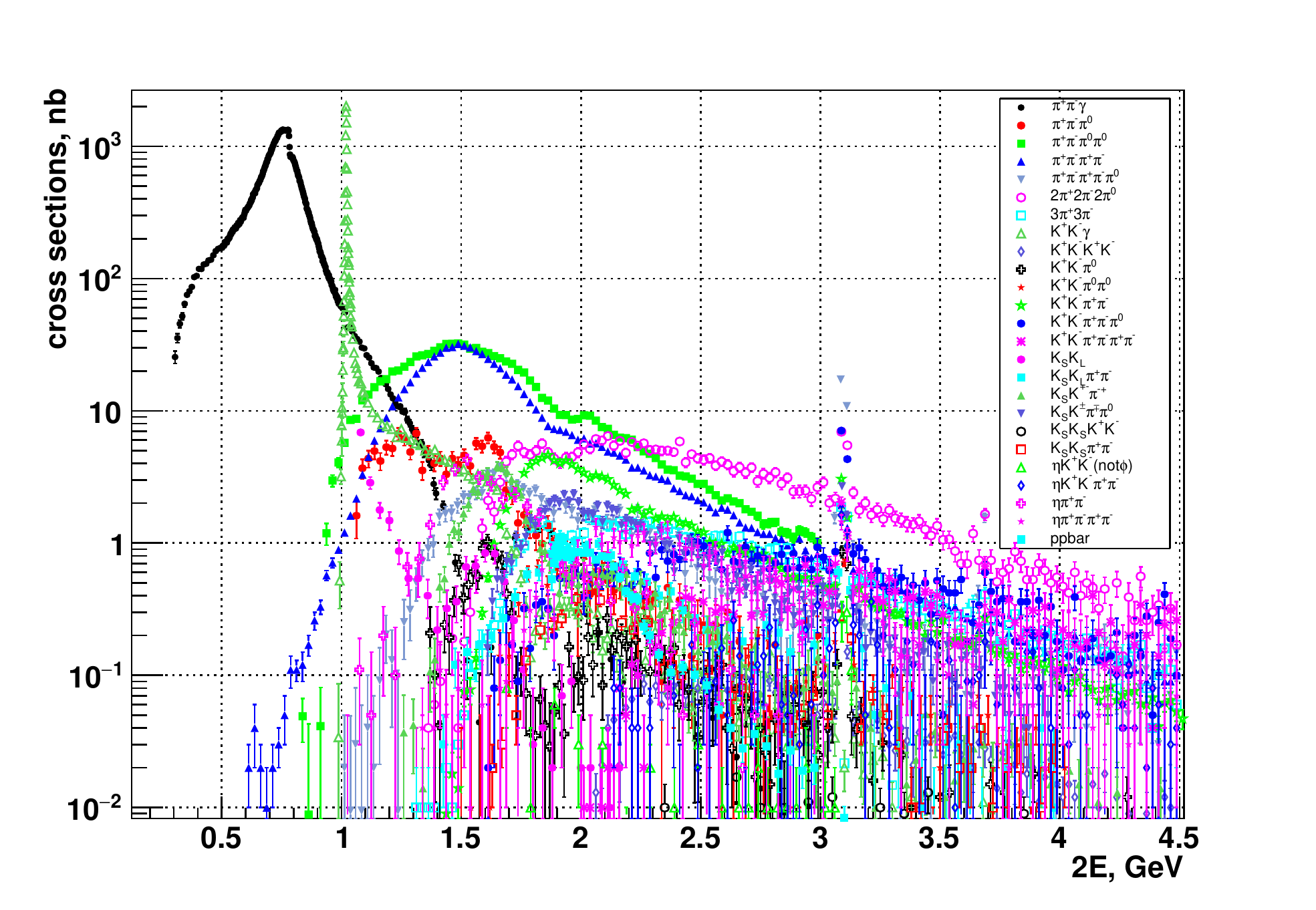}
}
\caption{Summary of the \babar\ measurements
(Courtesy of Fedor V. Ignatov, April 2016).
Beware that some channels have the charmonia contribution removed
while some others have not.
The $\pip\pim\piz\piz$ \cite{Druzhinin:2007cs}
and
 $\KS\Kp\pim \piz$  entries are preliminary.
The NLO measurements are denoted by an additional ``\g''.
\label{fig:summary:april2016}
}
\end{figure}

A summary of the \babar\ measurements is provided in 
Fig. \ref{fig:summary:april2016} and Table \ref{tab:compilation}.
The analyses of the 
$\pip\pim\piz\piz$ \cite{Druzhinin:2007cs}, of the 
$\pip\pim\piz$ \cite{Aubert:2004kj} and of the 
$\pip\pim\eta$ \cite{Aubert:2007ef}
channels are presently being updated with the full available
statistics: stay tuned.

\begin{table}[Htb]
 \small 
\caption{Summary of the \babar\ results on ISR production of exclusive
 hadronic final states
(The superseded results have been removed). 
Channels above the horizonal line have been mentioned in this paper.
\label{tab:compilation}
}


\begin{center} \footnotesize
\begin{tabular}{lcllllllll}
\hline 
\hline \noalign{\vskip3pt}
Channels & $\int {\cal L} \dd t$ (\invfb) & Method & Reference \\
\hline 
\hline \noalign{\vskip3pt}
$\KS\Kp\pim \piz$, $\KS\Kp\pim \eta$  & 454 & LO & preliminary 
\\
$\Kp\Km$ & 469 & LO, no tag  & \cite{Lees:2015iba}
\\
$\KS\KL$,  
$\KS\KL \pip\pim$, 
$\KS\KS \pip\pim$, 
$\KS\KS \Kp\Km $ & 469 & LO &  \cite{Lees:2014xsh}
\\
$\antiproton p $ & 454 & LO & \cite{Lees:2013ebn}  
\\
$\antiproton p $ & 469 & LO, no tag & \cite{Lees:2013uta} 
\\
$\Kp \Km$ & 232 & NLO & \cite{Lees:2013gzt} 
\\
$\pip\pim$ & 232 & NLO & \cite{Aubert:2009ad} \cite{Lees:2012cj} 
\\
\hline \noalign{\vskip3pt} 
$2(\pip\pim)$ & 454 & LO & \cite{Lees:2012cr} 
\\
$\Kp\Km \pip\pim$, 
$\Kp\Km \piz\piz$, 
$\Kp\Km \Kp\Km $ & 454 & LO & \cite{Lees:2011zi} 
\\
$\Kp \Km \eta$, 
$\Kp \Km \piz$, 
$\Kz \Kpm \pimp$ & 232 & LO & \cite{Aubert:2007ym} 
\\
$\pip\pim\piz\piz$ & 232 & LO & \cite{Druzhinin:2007cs} preliminary 
\\
$2(\pip\pim)\piz$ (including  $\pip\pim\eta$),
$2(\pip\pim)\eta$,
 & 232 & LO & \cite{Aubert:2007ef} 
\\
~ ~ ~ $\Kp \Km \pip \pim \piz$, 
$\Kp \Km \pip \pim \eta$
\\
$\Lambda \overline \Lambda $,
$\Lambda \Sigma^0$,
$\Sigma^0 \Sigma^0$
& 232 & LO & 
\cite{Aubert:2007uf} 
\\
$3(\pip\pim)$, 
$2(\pip \pim \piz)$, 
$\Kp \Km 2(\pip\pim)$ & 232 & LO & \cite{Aubert:2006jq} 
 \\
$\pip \pim \piz $ & 89 & LO & \cite{Aubert:2004kj} 
\end{tabular}
\end{center}
 
\end{table}

\begin{figure}[Htb]
\centerline{
\includegraphics[width=0.8\linewidth]{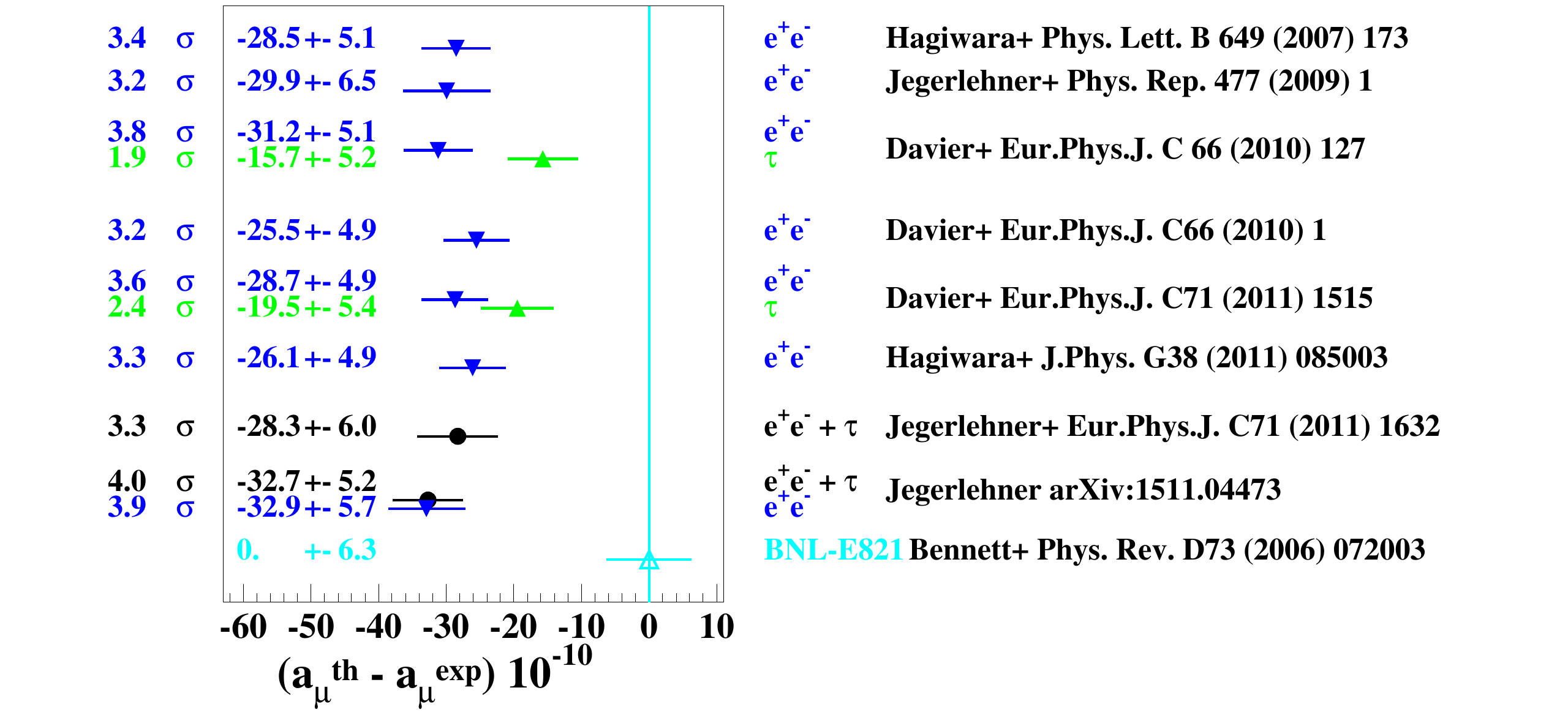}
}
\caption{
Recent predictions of the value of $a_\mu$ in chronological order
\cite{Hagiwara:2006jt,Jegerlehner:2009ry,Davier:2009ag,Davier:2009zi,Davier:2010nc,Hagiwara:2011af,Jegerlehner:2011ti,Jegerlehner:2015stw}, 
after the experimental value \cite{Bennett:2006fi} is subtracted.
\Blue Blue \Black: \epem-based; 
\Green Green \Black: $\tau$ spectral function-based; 
 Black: \epem and $\tau$ combinations.
\label{fig:combination}
}
\end{figure}

\clearpage

\section{What about $a_\mu$ then ?}

The time evolution of the prediction of
$a_\mu$ with the availability of experimental results of increasing
precision and with the development of combination techniques is shown in
Fig. \ref{fig:combination}.
\begin{itemize} 
\item
 After $\rho-\gamma$ mixing is taken into account, the discrepancy
 between the combinations based on \epem results and those based on
 the $\tau$ decay spectral functions \cite{Davier:2009ag}
 is resolved \cite{Jegerlehner:2011ti}.
\item
The discrepancy between the prediction and the measurement still sits
close to 3 -- 4 standard deviations.
\item 
Given that the precision of most of \babar\ measurements is now
dominated by the  contribution of the systematics, it will most likely be
difficult to achieve major improvements at a future super-$B$
factory.
\item 
Thanks to the high-precision results obtained up to the end of 2014,
the uncertainty on $a_\mu^{\VP}$ is now smaller than 
$4\times 10^{-10}$ \cite{Jegerlehner:2015stw}.
That work includes a NNLO correction for
$a_\mu^{\VP}$ \cite{Kurz:2014wya} and a NLO contribution to
$a_\mu^{\LbL}$ \cite{Colangelo:2014qya}.
Given the spread of the values predicted by the available models of
light-by-light scattering, the global uncertainty on $a_\mu^{\LbL}$
is of the same order of magnitude
\cite{Jegerlehner:2009ry,Jegerlehner:2015stw}.

\item 
Indeed, new measurements of $a_\mu$ at Fermilab
\cite{Venanzoni:2012qa} and at J-PARC \cite{Mibe:2011zz} are eagerly
awaited.
\end{itemize} 

\section{Acknowledgements}

Many thanks to the fellow BaBarians who helped me to prepare this talk
and to Fedor Ignatov who provided me with the \babar\ summary plot
(Fig. \ref{fig:summary:april2016}).

\end{document}